\documentclass[aps,superscriptaddress]{revtex4-1}
\usepackage{graphicx}
\usepackage{setspace}  
\usepackage{amsmath}
\usepackage{amssymb}  %%%  square symbol

\begin{document}             % End of document.

\title{Finite-size scaling for quantum criticality using the finite-element method}
\author{Edwin Antillon}
\affiliation{Departments of Physics,  Purdue University,  West Lafayette, IN 47907}
\author{Birgit Wehefritz-Kaufmann}
\affiliation{Departments of Physics,  Purdue University,  West Lafayette, IN 47907}
\affiliation{Departments of Mathematics,  Purdue University,  West Lafayette, IN 47907}
\author{Sabre Kais\footnote{kais@purdue.edu}}
\affiliation{Departments of Physics,  Purdue University,  West Lafayette, IN 47907}
\affiliation{Department of Chemistry, Purdue University, West Lafayette, IN 47907}

\begin{abstract}
Finite size scaling for the Schr\"{o}dinger equation is a systematic approach
to calculate the quantum critical parameters for a given Hamiltonian. This
approach has been shown to give very accurate results for critical parameters 
by using a systematic expansion with global basis-type functions. Recently, the finite element method 
was shown to be a powerful numerical method for \emph{ab initio} electronic structure calculations 
with a variable real-space resolution. In this work, we demonstrate how to obtain quantum critical 
parameters by combining the finite element method (FEM) with finite size scaling (FSS) using
different ab initio approximations and exact formulations.
The critical parameters could be atomic nuclear charges, internuclear distances, electron density, disorder,
lattice structure, and external fields  for stability of atomic, molecular systems and quantum
phase transitions of extended systems. 
To illustrate the effectiveness of this approach we provide detailed 
calculations of applying FEM to approximate solutions for the two-electron 
atom with varying nuclear charge; these include Hartree-Fock, density functional theory under
the local density approximation, 
and an ``exact'' formulation using FEM. We then use the FSS approach to determine its critical
nuclear charge for stability; here, the size of the system is related to the number of
elements used in the calculations. Results prove to be in good agreement with previous Slater-basis set
calculations and demonstrate that it is possible to combine finite size scaling with the finite-element method 
by using ab initio calculations to obtain quantum critical parameters. The combined approach provides a promising 
first-principles approach to describe quantum phase transitions for materials and extended systems. 
\end{abstract}

\maketitle

\section{Introduction}

The theory of finite size scaling (FSS) in statistical mechanics can provide us with numerical methods \cite{fisher,widom,barber,privman,cardy,nightingale1,Peter1,Peter2} capable of obtaining accurate results for infinite systems by studying the corresponding small systems. In the past, basis set approximation to many-body systems have been combined with FSS to obtain critical parameters, where calculations for these systems were done by expanding the wave function in Slater-type basis sets or Gaussian-type basis functions. However, a generalization of larger atomic and molecular systems proved itself difficult using these type of functions \cite{MSK,PSK}.

More recently, we have combined FSS with the finite element method (FEM) in order 
to obtain critical parameters in a potential that admits a finite number of bound states 
and we obtained excellent results by using both finite-element and Slater-type basis functions \cite{AMWK}.
For this potential we also established that the finite-difference (FD) method can be combined with FSS.
The finite difference method, however, is not a variational method since it approximates the operators 
(i.e. the derivative) and therefore can result in errors of either sign \cite{TT2}.
As a result, we observed non monotonic behavior of the calculated critical parameters with respect to the number (N)
of grid points used \cite{Daniel}. Nevertheless, for a large enough number of grid points N
one can recover the correct asymptotic behavior in the large N limit. 
For this particular example we used FD to fourth order and in FEM we used $C^{1}$-continuous basis.

In this paper, we present a method to combine FEM with FSS for electronic-structure calculations near 
regions of criticality. 
We note that \emph{any} method where the wavefunction can be expanded in a complete basis (or set of grid points)
should also work, provided that in the infinite basis limit ($N\rightarrow \infty$) it becomes 
arbitrarily close to the the exact wavefunction.
The main advantage of finite element methods, over other global basis methods, 
resides in the fact that matrices obtained by these numerical methods are often banded and tend to be 
sparse since variables are not coupled over arbitrarily large distances. This
is ideal for describing extended systems since it facilitates parallelization \cite{pask}. Moreover, 
due to the polynomial nature of the basis set most integrals involved can be evaluated easily and accurately
with a better convergence with respect to the basis set size.

We examine the two electron system with different levels of approximations to the ground state energy,
by solving the system using FEM both \emph{exactly} and using \emph{mean-field equations} and
demonstrate that this formalism is general and that it has potential for more complex systems by solving
Hartree-Fock equations or Kohn-Sham equations in density-functional theory.
%%%%%%%% edit
The paper is  organized as follows: 
In Sec. (II) we discuss FEM formalism,
in Sec. (III) we discuss the FSS approach to quantum systems, 
in Sec. (III) and (IV) we combine FSS with FEM to compute calculations of critical parameters for two electron atoms, 
,and, lastly in Sec. (V) we discuss the results and applications to extended systems. 
%%%%%%%%
\section{Finite Element Method in Electronic Structure Calculations}
Computational methods in quantum chemistry typically make use of Slater or Gaussian basis sets to approximate solutions 
in atomic or molecular systems. The advantages of these methods are that some of the integrals involved in the 
calculations are analytical and very accurate results can be obtained for these integrals. 
However, there are problems in extrapolating to the complete basis limit.
Plane wave methods, having a complete basis, present some difficulties if non-periodic boundary conditions 
are to be used such as in molecule calculations. A common trait of all these methods is that they employ global basis,
which can result in some difficulties for the formulation of large problems. First, the 
matrix formulation becomes dense using global basis and therefore ill-conditioned for parallelism \cite{pask}. 
While for some complex objects, convergence issues may arise if there are large variations of potential terms 
in small regions which cannot be captured by the traditional Gaussian basis set that are centered on the nuclei.

The FEM presents an alternative solution to these problems. This method
uses a variational method to solve differential equations by discretizing a continuous solution to a set of values in sub-domains, called elements, and employs a \emph{local} basis expansion in terms of polynomials with variable real-space resolution, allowing for convergence to be controlled systematically. 
The locality of the bases in real space results in sparse and banded matrices, where the number of operations for 
matrix-vector multiplication can be reduced from $O(N~Log~N)$ for dense matrices to $O(N)$ for sparse matrices, where {\bf N} is the dimension of the matrix \cite{pask}. 
Some of the first applications of FEM in engineering date back to 1950,
in problems such as elasticity and structural analysis in civil engineering and aeronautical engineering, respectively.
In the 1970s, some applications to quantum mechanical problems appeared, but were limited to
small systems due to the storage limitation \cite{FEM-Book1,FEM-Book2}
Recently, advances in high-performance computing make FEM a viable alternative to the traditional approaches of electronic-structure calculations \cite{FEM-Review1,FEM-Review2}.
The sparsity of the global matrices resulting from the FEM makes the parallel formulation more affordable. Alizadean \emph{et al}, have recently applied a \emph{divide-and-conquer} method to the FEM-Hartree-Fock approach for electronic 
calculations showing a facilitation on parallelization and reduced scaling for larger systems \cite{AHM}.

We briefly describe the FEM formulation for a given Hamiltonian to be used in later sections.
Let us consider the following one-particle Hamiltonian:
\begin{equation}\label{eq:H}
  -\frac{1}{2}\nabla^2 \Psi_i + V \Psi_i = \epsilon_i \Psi_i
\end{equation}
where $\epsilon_i$ and $\Psi_i$ are the energy eigenvalue and eigenvector respectively,
 and V is an arbitrary potential that 
will represent a sum of terms such as the electron-nuclear potential ($V_{en}$), Hartree potential ($V_H$) and exchange-correlation potentials ($V_{xc}$).
We can reformulate this problem into a \emph{weak form} by taking an inner product of an arbitrary ``\emph{test} function'' $\Phi$ with Eq (\ref{eq:H}):
\begin{equation}\label{eq:weak}
  \int_\Omega \Phi[-\frac{1}{2}\nabla^2 + V - \epsilon_i]~\Psi_i d\Omega = 0 
\end{equation}
This relaxes the problem into a variational form where instead of finding an exact solution in the domain $\Omega$, we find a solution that satisfies Eq.(\ref{eq:H}) in an average sense everywhere. We then arrive at an equivalent equation by integrating by parts (in real space):
\begin{equation}\label{eq:byparts}
  \int_\Omega \frac{1}{2}\nabla\Phi(r) \cdot \nabla\Psi_i(r)~d\Omega - \int_\Gamma \frac{1}{2} \Phi(r) \nabla\Psi_i(r) \cdot \hat{n} ~d\Gamma  + \int_\Omega \Phi(r) (V - \epsilon_i)~\Psi_i(r) d\Omega = 0
\end{equation}
where $\Gamma$ is a boundary term and $\hat{n}$ is the outward unit normal at {\bf{r}}. 
In the absence of external fluxes we can employ Neumann boundary conditions 
[$\nabla \Psi_i(r)\cdot \hat{n}= f(r)~~r\in\Gamma $]. Furthermore, in radial coordinates ($d\Omega=r^2 dr$)
we can write the surface term in the following form: 
\begin{equation}
  \int_\Gamma \frac{1}{2} \Phi(r) \nabla\Psi_i(r) \cdot \hat{n} ~d\Gamma  =  [\frac{1}{2} \Phi(r)~f(r)~r^2 ]^{r_{cut}}_{0} 
\end{equation}

This surface term vanishes trivially at $r=0$, while for $r=r_c$, it can be removed if we choose a test function 
$\Phi(r)$ that vanishes at $r=r_{cut}$. Since Eq (\ref{eq:byparts}) is satisfied for any $\Phi(r)$, we can choose
$\Phi=\Psi^*$ and in this manner it reduces the problem to an eigenvalue problem. The wavefunction $\Psi_i(r)$ can be then discretized using the  \emph{Garlekin} approximation \cite{FEM-Book1,FEM-Book2}:
\begin{equation}\label{eq:Garlekin}
  |\Psi(r) \rangle \approx {\bf{N}^{T}} |{\bf{\psi}}\rangle = \sum_i N_i(r)  |\psi_i \rangle
\end{equation}
where $N_i(r)$ are piecewise-linear $C^0$ continuous functions (i.e. continuous but not necessarily smooth) and $|\psi_i \rangle$ are the nodal degrees of freedom. Smoother bases can be constructed by using higher order polynomial ($C^n$ continuous, $n\geq1$) and by extending the nodal space to include the derivative of the wavefunction at the node, viz,   $|\Psi \rangle = \sum_i N_i(r) |\psi_i \rangle + \partial_x N_i(r) | \partial_x \psi_i \rangle $; this ensues continuity of the wavefunction between the nodes. In the nodal basis, the generalized eigenvalue problem reads

\begin{equation}\label{eq:genH}
  \sum_{ij} \langle \psi_i | H_{ij} |\psi_j \rangle =  \epsilon \sum_{ij} \langle \psi_i | S_{ij} |\psi_j  \rangle
\end{equation}
where the above terms can be written in terms of local basis representation, viz
$$  H_{ij} = \int_\Omega [\frac{1}{2} \nabla N_i \cdot \nabla N_j^{T} +  V N_i N_j^{T}] d\Omega $$
$$  S_{ij} = \int_\Omega N_i N_j^{T} d\Omega $$

The finite element method takes advantage of the strict locality of its basis to yield matrices that are sparse 
and structured and due to the polynomial nature of the basis most integrals can be easily evaluated, 
as in global basis approaches. Moreover, since the method is variational, all errors are of the same sign, leading to monotonic convergence often from below \cite{pask2}.

\section{Finite Size Scaling}
We now discuss the role of finite size scaling in systems exhibiting critical behavior.
Phase transitions are associated with singularities of the free energy that occur only in the thermodynamic limit \cite{yanglee1,yanglee2}.
Near criticality, fluctuations in the system become correlated over distances of the order of the correlation length $\xi$, while for 
second order phase transitions, this correlation length $\xi$ diverges.

The theory of finite size scaling can be used to extract universal parameters describing the transition by studying systems of finite size \cite{widom,barber,privman,cardy,nightingale1,Peter1,Peter2}. 
Any physical quantity ($P$) that would diverge in the thermodynamic limit, becomes now ``rounded off'' in the finite systems (of size L) by a scaling function 
$f(x)$ \cite{cardy,nightingale1}
\begin{equation}\label{eq:fss}
  \frac{P_L(t)}{P_{\infty}} = f(\frac{L}{\xi_{\infty}(t)}),
\end{equation}
where $t= (T -T_c)/T_c$ is a reduced critical parameter describing the phase transition and $\xi_\infty$ is 
the correlation length of the system in the thermodynamic limit.

While classical phase transitions are governed by the fluctuation of temperature 
that tend to minimize the free energy, at zero temperature, these fluctuations are 
driven by Heisenberg's \emph{uncertainty principle} \cite{sachdev}.
A signature of a phase transition is \emph{non-analyticity} of the ground state energy
$E(\lambda)$ with respect to a parameter $\lambda$ in the Hamiltonian. The non-analyticity can
be the result of avoided crossing, or actual level crossing.
In the first case, the configuration of the system that minimizes the energy maintains 
the energy gap ($\Delta E_\lambda$) between the first excited energy state and the ground state as the
parameter $\lambda$ is varied. In the latter case, the energy levels cross so that 
the energy is now minimized by a configuration that previously corresponded 
to the first excited state of the system. 

To pin down the transition point involves computing the energy accurately near the critical point. However,
this can be quite a challenge since most real systems lack an analytical solution to the ground state energy
and approximation methods, such as perturbation theory or mean-field methods (Hartree-Fock , DFT) which are 
viable options in stable systems, in general are not well defined or become ill-conditioned near critical points.

This is where the theory of finite size scaling can be recast in the context of the Hilbert space.
Any given approximation to the ground state energy will result in a ``rounding'' effect 
from the ``true'' critical behavior, since any approximation to a wavefunction can be expanded in some basis and truncated to some order N; that is,

\begin{equation}
  | \Psi \rangle = \sum_i^N a_i | \phi_i \rangle
\end{equation}

FSS has been used in quantum systems \cite{neirotti0,serra2,kais,snk1,snk2,nsk,qicun,kais1,review,adv}. 
In this approach, the finite size corresponds not to the \emph{spatial} dimension but to
the \emph{number of elements} ({\bf{N}}) in a complete basis set used to expand the exact
eigenfunction of a given Hamiltonian \cite{dipole,quadrupole,qicun1,qicun2,Ferron-Last}.
In the variational approach singularities in different mean values will occur only in the limit of infinite
number basis function, viz.,
\begin{equation}
  \left< O \right>_\lambda^{\infty} \sim  (\lambda - \lambda_c)^{\mu_o} 
\end{equation}

The FSS \emph{ansatz} in the Hilbert space takes the following form (for expectation values of 
operators acting on the Hilbert space). There is a scaling function $F(x)$ such that 
\begin{equation}
  \label{eq:fss_QM}
  \langle O \rangle_\lambda^{(N)} \sim  \langle O \rangle_\lambda^\infty F_O  (N|\lambda-\lambda_c|^\nu)
\end{equation}
where $\lambda$ is a critical parameter in the Hamiltonian, 
$O$ is a quantum mechanical operator near criticality, and $F_O$ is a scaling function for each different operator 
but with a unique scaling exponent. At the critical point, the expectation value is related to $N$ as a
power law, $\left<O\right>^{(N)} \sim N^{-\mu_o/\nu}$. In this manner the scaling function $F_O$ will remove the divergence as it is computed in the finite basis.

The asymptotic behavior of operators near the critical point, in this case the ground state energy, can be expressed 
using critical exponents similarly to classical phase transitions, viz

\begin{equation}
  \label{eqn:gap}
  \Delta E_\lambda \sim (\lambda - \lambda_c)^\alpha
\end{equation}
where $\alpha$ is a universal critical exponent. In addition to the vanishing
energy scale, a diverging length scale also occurs which is associated with
the exponential decay of equal-time correlations in the ground state of the system \cite{sachdev}.
For our case, we focus on the behavior of the wavefunction at threshold, where the correlation length is defined by:
\begin{equation}
  \xi  \sim |\lambda-\lambda_c|^{-\nu} 
\end{equation}
In order to apply finite size scaling method to quantum system, we assume a Hamiltonian of the form
\begin{equation}
  H  =  H_o + V_\lambda
\end{equation}
where $\lambda$ is the coupling parameter showing a phase transition at $\lambda=\lambda_c$.
For potentials linear in $\lambda $ ( $V_\lambda    = \lambda V $), Simon \cite{simon} showed that the critical
exponent is equal to one if and only if the Hamiltonian $H(\lambda_c)$ has a normalizable eigenfunction
with eigenvalue equal to zero. The existence or absence of a bound state at the
critical point is related to the type of singularity in the energy. Using statistical
mechanics terminology, we can associate a``first order phase transition'' with the
existence of a normalizable eigenfunction at the critical point. The absence of such 
a function could be related to a ``continuous phase transition.''
To obtain the critical parameters, we define the following function that will remove the N-dependence at a critical point:
\begin{equation}
\triangle_O(\lambda;N,N')=\frac{\ln(\left<O\right>_
\lambda^{N}/\left<O\right>_\lambda^{N'})} {\ln(N'/N)}
\label{fourteen}
\end{equation}
For the energy operator $O=H$, and using the customary $\alpha$ Greek letter for the corresponding exponent $\mu_o$ we have
\begin{equation}
\triangle_H(\lambda_c;N,N')=\frac{\alpha}{\nu}
\label{alphanu}
\end{equation}
To arrive at an expression that isolates the critical exponent $\alpha$ from numerical calculations, we make 
use of the \emph{Hellmann-Feynman} theorem \cite{landau}
\begin{equation}
 \frac{d E_{\lambda}}{d\lambda} = \left< \frac{d H_{\lambda}}{d\lambda} \right> 
\end{equation}
At this point it becomes convenient to define another function that will be used in the following sections.
\begin{equation}\label{eq:gammafunc}
\boxed{
\Gamma_\alpha(\lambda,N,N')=\frac{\triangle_H(\lambda;N,N')}{\triangle_H(\lambda;N,N')-\triangle_{\frac{\partial
      H_\lambda}{\partial \lambda}}(\lambda;N,N')}
}
\end{equation}
At the critical point this function is independent of $N$ and $N'$ and takes
the value of $\alpha$. Namely, for $\lambda=\lambda_c$ and any
values of $N$ and $N'$ we have
\begin{equation}
\Gamma_\alpha(\lambda_c,N,N')=\alpha
\end{equation}
\section{Implementation }
In the following, we illustrate the combined approach (FEM+FSS) for the two-electron atom, but emphasize that the method is general and can be applied to more complex atomic and molecular systems. We seek the minimum charge (\emph{critical charge}) needed to bind both electrons to the nucleus with charge Z.
It is well known that ignoring the electron-electron energy correlation will not predict the stability of $H^-$ 
\cite{chandr}, since the \emph{critical charge} obtained by such assumption is $Z_c \approx 1.02$.
Stillinger (1974) \cite{Stillinger} took into account the e-e correlation by using a 
non-linear wavefunction $e^{-a r_1-b r_2} + e^{-a r_1-b r_2}$  in order to find the minimum energy as the nuclear charge Z was varied. He found a critical charge around $Z_c \approx .9537$. Baker \emph{et al} (1990) \cite{Baker} 
performed a thorough 401-order perturbation calculation to resolve the controversy for the critical charge; they obtained
a critical charge of $Z_c \approx .91103$. This is the value we use as reference. 
In the following sections we illustrate the mean-field equations and exact formulations for the
ground state of the two-electron system, while in the next section we apply FSS to these approximations. 
Let us start with the Hamiltonian for the two-electron atoms; it is given by
\begin{eqnarray}
  \label{eqn:Hamiltonian_0}
  H & = & -\frac{1}{2}\nabla^2_1 -\frac{1}{2}\nabla^2_2 + \frac{e^2}{|r_1 -r_2|} - (\frac{Z}{r_1} + \frac{Z}{r_2}),
\end{eqnarray}
where $r_1$ and $r_2$ are the coordinates of the two electrons referenced from the nuclei. 
We use Hartree units ($\hbar = m = e = 1$) throughout in our calculations.

\underline{\textsc{Hartree Fock (HF) Approximation}}: \\ 

In the Hartree approximation the total wavefunction is approximated as a product of single particle wavefunctions is $\Psi(r_1,r_2) = \psi_1(r_1) \psi_2(r_2) $. We can arrive at a self-consistent form by multiplying the Hamiltonian by the complex conjugate of the aforementioned wavefunction and 
integrating one coordinate out, we arrive at
\begin{equation}
  \label{eqn:Hartree}
  [-\frac{1}{2}\nabla^2_1 - \frac{Z}{r_1} + \int d^3r_2 \frac{1}{|r_1 -r_2|} |\psi_2(r_2)|^2 ]\psi_1(r_1) = \epsilon \psi_1(r_1)
\end{equation}
Using the FEM (see Sec. II) this expression can be reduced to a generalized eigenvalue problem $ H_{kl}  \psi_l =  \epsilon  S_{kl} \psi_l $. The Hamiltonian $H_{kl}$ is written in terms of the single electron energy term $h_{kl}$ (i.e. $-\frac{1}{2}\nabla^2 -\frac{Z}{r}$) and a \emph{direct} coulomb term $J_{kl}$ defined below, viz
\begin{equation}
  H_{kl} = \int [\overbrace{\frac{1}{2} \nabla N_k* \nabla N_l^{T} + \frac{1}{r} N_k N_l^{T}}^{h_{kl}}  +  \overbrace{\sum_{ij} Q_{klij} \psi_i \psi_j}^{J_{kl}}] d^3r_1
\end{equation}
$$  Q_{klij} \equiv N_k(r_1)^* N_l(r_1) \int  \frac{1}{|r_1 -r_2|}  N_i(r_2)^* N_j(r_2) d^3r_2$$
$$  S_{kl} = \int N_k N_l^{T} d^3r_1 $$

Since electrons are fermions, the product of the single particle states needs to be antisymmetric with respect to exchange
of the coordinates on the electrons. This can be achieved by writing the wave function as a Slater determinant \cite{thijssen}:
\begin{equation}
 \Psi(r_1,r_2) =
 \begin{vmatrix}
   \psi_1(r_1) & \psi_2(r_1)  \\
   \psi_1(r_2) & \psi_2(r_2)  \\
 \end{vmatrix}
\end{equation}

Searching for the best Slater determinant self-consistently in the Hamiltonian [Eq.\ref{eqn:Hamiltonian_0}] leads to
the Hatree-Fock equations \cite{thijssen,Schw,GG}.
\begin{equation}
  \label{eqn:HF}
  \int \psi^*_1(r_1) (-\frac{1}{2}\nabla^2_1 - \frac{Z}{r_1}-\epsilon)\psi_1(r_1) d^3r_1 + \frac{1}{2}\int \int \{|\psi_1(r_1)|^2 \frac{1}{|r_1 -r_2|} |\psi_2(r_2)|^2 - \psi^*_1(r_2) \psi^*_2(r_2) \frac{1}{|r_1 -r_2|} \psi_1(r_1)\psi_2(r_1)\}d^3r_1 d^3r_2 = 0
\end{equation}
where the additional term on the right is known as the \emph{exchange} Coulomb term, denoted by  $K_{kl} $, 
and tends to repel states of the same (parallel) spin. 
We can write this term in the local basis form by looking at the direct Coulomb term and 
noting that the indices $(i,j) \in \psi_2$ 
and similarly indices $(k,l) \in \psi_1$. The operator $K_{kl} $ is obtained from  $J_{kl} $  by switching indices ($ j \leftrightarrow l$), which effectively switches the coordinates $r_1$  and $r_2$ on two of the wavefunctions and makes this operator non-local.

The ground state energy (for $N_e$ electrons) is determined by solving for the ground state eigenfunction 
$|\psi_g \rangle$ for the eigenvalue problem (Eq.\ref{eqn:HF}) and computing the expectation value below (to
avoid double-counting terms),
\begin{equation}
  E_{HF} =  \sum_j^{N_e} \langle \psi_g | h_j -\frac{1}{2}  (J_j - K_j)  |\psi_g \rangle 
\end{equation}

Note that for our case, the exchange term does not arise since the ground state of the helium atom 
does not have parallel spins. In order to improve upon the Hartree-Fock approximation we can 
take into account the correlation energy from an approximation to the free electron gas due to Wigner. Here the ``correlation'' potential can be obtained from Refs. \cite{wigner1,wigner2}. Note that the references cited use Rydbergs units, and reference \cite{wigner1} additionally uses $e^2=2, \hbar^/2m=1$. In Hartrees
the correlation potential is given by:

\begin{equation}
  V_c(r) = \frac{0.29}{5.1 + r_s(r)}
\end{equation}
where $r_s$ is the local density parameter for the atom, that is
\begin{eqnarray}
  \rho(r) &=& \frac{1}{\frac{4}{3}\pi r_s^3} \\
  \rho(r) &=& \rho(r)_{\uparrow} + \rho(r)_{\downarrow} 
\end{eqnarray}
The last expression follows for helium since in the ground state the two electrons have opposite spin, but share the 
same spatial orbital. In this manner the correlation potential is included self-consistently and the density is obtained from the sum of the densities 
for each electron i.e.,  $\rho(r) = 2 |\psi(r)|^2$. Finally, the correlation energy is obtained as
$$  E_{corr} = \langle \psi_g | V_c(r) | \psi_g \rangle $$
and the value for the \emph{total energy} is given by:
\begin{equation}
E_{total} = E_{HF} + E_{corr}
\end{equation} \\

\underline{\textsc{Density Functional Theory}} \\ 

In the Hartree-Fock approximation, the many-body wavefunction is written in the form 
of a Slater determinant while the correlation energy is treated separately. This
can be done for small systems, but for larger systems this is not an easy task. 
The theory of Kohn and Sham developed in 1965 simplifies both the exchange and correlation 
effect of electrons by reducing their effect to a effective potential $V_{KS}$ that is
{\it in principle} exact. In their seminal paper \cite{KS}, they showed that to determine the 
ground state energy it is sufficient to determine the ground state electron density $\rho(r)$.
By constructing a simple system of {\it non-interacting} electrons which yields 
the same density as the ``real'' fully interacting system, the ground state energy can be obtained. 

The {\it Kohn-Sham equations} for two electrons are:
\begin{equation}
  \label{eqn:KS}
  [-\frac{1}{2}\nabla^2_1 - \frac{Z}{r_1} + \int d^3r_2 \frac{1}{|r_1 -r_2|} \rho(r_2) + V_{xc}(\rho) ] \psi_i(r_1) = \epsilon \psi_i(r_1)
\end{equation}
These equations are self-consistent in the density and differ from the Hartree-Fock equations only in the exchange-correlation potential  $V_{xc}[\rho]$
which is a function of the total local density: $  \rho(r) = \sum_i^N |\psi_i(r)|^2 $ 
\begin{equation}
  \label{eqn:der}
  V_{xc}[\rho(r)] = \frac{\delta}{\delta \rho(r)} E_{xc}[\rho(r)]
\end{equation}
While the exact form of the  exchange-correlation functional $E_{xc}[\rho]$ is not known, the 
simplest approximation is the local density approximation (LDA)
\begin{equation}
  E^{LDA}_{xc}[\rho(r)] =   E_{x}[\rho(r)] +  E_{c}[\rho(r)] 
\end{equation}
\begin{equation}
  E_{x}[\rho(r)] =  -(\frac{3}{4})(\frac{3}{\pi})^{1/3} \int \rho(r)^{4/3} d^3r 
\end{equation}
\begin{equation}
  E_{c}[\rho(r)] =  \int \frac{-0.29 \rho(r)}{5.1+r_s(r)} d^3r
\end{equation}
It follows from Eq.(\ref{eqn:der}) that the potentials are given by:
\begin{equation}
  V_{x}[\rho(r)] = -(\frac{3}{\pi})^{1/3} \rho(r)^{1/3}
\end{equation}
\begin{equation}
  V_{c}[\rho(r)] =  \frac{-0.29 }{5.1+r_s(r)} -\frac{1}{3}\frac{0.29 r_s(r) }{(5.1+r_s(r))^2} 
\end{equation}
and the ground state energy of the system is obtained from \cite{thijssen}:
\begin{equation}
  E_{DFT}=  \sum_i^{N_e} \epsilon_i - \frac{1}{2} \int \int d^3r_2 d^3r_1 \rho(r_1) \frac{1}{|r_1 -r_2|} \rho(r_2) + E_{xc}[\rho] 
  - \int d^3r V_{xc}[\rho(r)] 
\end{equation}

\begin{table}\small
  \caption{
    Comparison of various components to the Kohn-Sham and Hartree-Fock energies in Hartrees for helium. 
    The reference values ({\bf Perdue-Wang, Exact}) are correct to all digits quoted \cite{UG,PW}, while 
    the accuracy of the {\bf FEM} estimates on $E_{tot}$ is expected to be on the order of two digits (0.01) for N=100 
  and three digits (0.003) for N=200 for both {\bf LDA} and {\bf Total Energy}, refer to the discussion on the text for details.} 
  \centering \bf
  \begin{tabular*}{0.95\textwidth}{@{\extracolsep{\fill}} | l  |c c c | c c c | }
    \hline\hline
    & \multicolumn{3}{c|}{\large{LDA}} &   \multicolumn{3}{c|}{\large{Total Energy}} \\
    &FEM      &FEM       &Perdue-Wang   &FEM     &FEM       &Exact\\
    &(N=100)  &(N=200)   &\cite{UG,PW}    &(N=100) &(N=200)   &\cite{UG}\\
    \hline
    $E_{tot} = E_{Kin} + E_{en} + E_H + E_{xc}$  &  -2.813 661  &  -2.821 852  & -2.834 455  &  -2.886 297 &  -2.904 925 & -2.903 724 \\
    $E_{Kin}$                                    &   2.703 352  &   2.726 064  &  2.767 389  &   2.806 848 &   2.855 856 &  2.867 082 \\
    $E_{en}$                                     &  -6.538 681  &  -6.571 185  & -6.624 884  &  -6.664 319 &  -6.737 783 & -6.753 267 \\
    $E_{H}$                                      &   1.973 543  &   1.978 456  &  1.995 861  &   1.020 329 &   1.026 754 &  1.024 568 \\
    $E_{x}$                                      &  -0.850 814  &  -0.854 086  & -0.861 740  &   0.000 000 &   0.000 000 &  0.000 000 \\
    $E_{c}$                                      &  -0.101 061  &  -0.101 092  & -0.111 080  &  -0.049 135 &  -0.049 814 & -0.042 107 \\
    $\epsilon$                                   &  -0.563 621  &  -0.565 805  & -0.570 256  &  -0.957 555 &  -0.963 991 & -0.903 724 \\
    \hline
    \hline
  \end{tabular*}
  \label{table:Energies}
\end{table} 

The Kohn-Sham equations are in general simpler than HF equations, while including correlation effects beyond the HF approximation. Furthermore their computing time grows as a power of the number of the N-electrons, whereas the exact solution
for the N-electron Schrodinger equations is exponential in N \cite{GG}.\\

Let us comment on the difference between the two approaches, 
even though both Hartree-Fock and Kohn-Sham equations have similar potential terms:  
$  V_{KIN}, V_{en}, V_{H}, V_{xc} $ [see Eqs.(\ref{eqn:HF}) and (\ref{eqn:KS})].
In HF the energy is calculated using the wavefunction explicitly, whereas in DFT the energy is a functional of the density. 
For example, for the correlation energy we have in the HF approximation:
$$  E_c^{HF} = \int  |\psi(r)|^2 V_c(r) dr^3 $$
while for the DFT approximation we have:
$$  E_c^{DFT} = \int \rho(r) V_c[\rho(r)] dr^3 $$

Care should be exercised when computing the LDA exchange potential since this term is non-linear in the density
and as a result the normalization does not trivially cancel out when solving the eigenvalue problem. 
For example, in this work we use the normalization $\int  |\psi(r)|^2 r^2 dr = 1 $ so that a factor of $\sqrt(4\pi)$ is absorbed into the wavefunction $\psi(r) \rightarrow \psi(r)/\sqrt(4\pi)$. This in turn changes the density to $\rho = 2 |\psi(r)|^2 \rightarrow 2 \psi(r)^2/4\pi$, and therefore the exchange-potential in LDA becomes

$$  V_{x}[\rho(r)] = -(\frac{3}{\pi})^{1/3} \rho(r)^{1/3} \rightarrow -( \frac{3 \psi(r)^2}{2\pi^2})^{1/3} $$

In order to satisfy the boundary conditions $\psi(r)\rightarrow 0$ as $r\rightarrow \infty$, we use a uniform mesh
with a cutoff radius at $r_{cut} = 10~a.u$. The radial equations where solved using $C^0$-continuous polynomials, that is, $N_{L \rightarrow R}(x) = x $ and $N_{L \leftarrow R}(x) = 1-x $; where the function $N_{L \rightarrow R}(x)$ interpolates the wavefunction from the left node to the right node and similarly for $N_{L \leftarrow R}(x)$.
We solved all the integrals numerically using 10-point Gaussian quadratures and in order to solve the generalized eigenvalue problem we have used a fortran subroutine EWEVGE \cite{EWEVGE}.

In Table \ref{table:Energies} we show the energy components of the Kohn-Sham equations under LDA approximation 
and the Hartree-Fock equations using the Wigner correlation potential for the helium atom.
We note that the reference values for the Total Energy (HF+Wigner correlation) are exact, therefore even 
in the infinite basis limit the expression will not converge to the value shown. 
We also point out that the Perdue-Wang values on the correlation energy are obtained by using the
random-phase approximation (RPA) for the free electron gas \cite{PW}. 
Our convergence value is be limited by how well we can approximate this value. 
We have checked our level of convergence by increasing the number of elements to N=1000 in a linear mesh of size 
$r_{cut} = 10~a.u$ with $C^0$-continuous basis (linear polynomials). 
We obtain a total energy of $E_{total}^{FEM} = −2.824596$ and correlation energy of $E_{c}^{FEM} = -0.101103$, and when 
compared to the values by Perdue-Wang (see Table \ref{table:Energies} ), the errors are $\delta E_{total} = 0.009859$ and $\delta E_{c} = 0.009977$ respectively. 
We see that the error in the total energy is dominated by the correlation term arising from using Wigner correlation rather than RPA description. However, taking this difference into account we can get an estimate for the error 
of all other terms, viz  $\delta E_{Kin} + \delta E_{en} + \delta E_H + \delta E_x \approx \delta E_{total} - \delta E_{c} =
0.000118$. In FEM, the error on $L^2$ integrals such as the energy is $O(h^{(p+1)})$, 
where h is the element width and p is the order of the polynomial used. Thus, for our example using N=1000 
in a uniform grid size with $r_{cut} = 10$ a.u. we have $h = 0.01$ , and since we employ a linear basis, the error we 
expect is $O(h^{(1+1)}) = 0.0001$. In this manner, if we take into account the intrinsic error due to the different 
energy-correlation terms we can recover the level of convergence expected by using FEM.\\

\underline{\textsc{Exact Formulation}}\\

%%%%%%%%%%%%%%%%%%%%%%%%%%%%%%%%%%%%%%%%%%%%%%%%%%

%\par\vspace{5mm}
\begin{figure}[t]
  \centering
  \begin{tabular}{cc}
    \includegraphics[scale=.5]{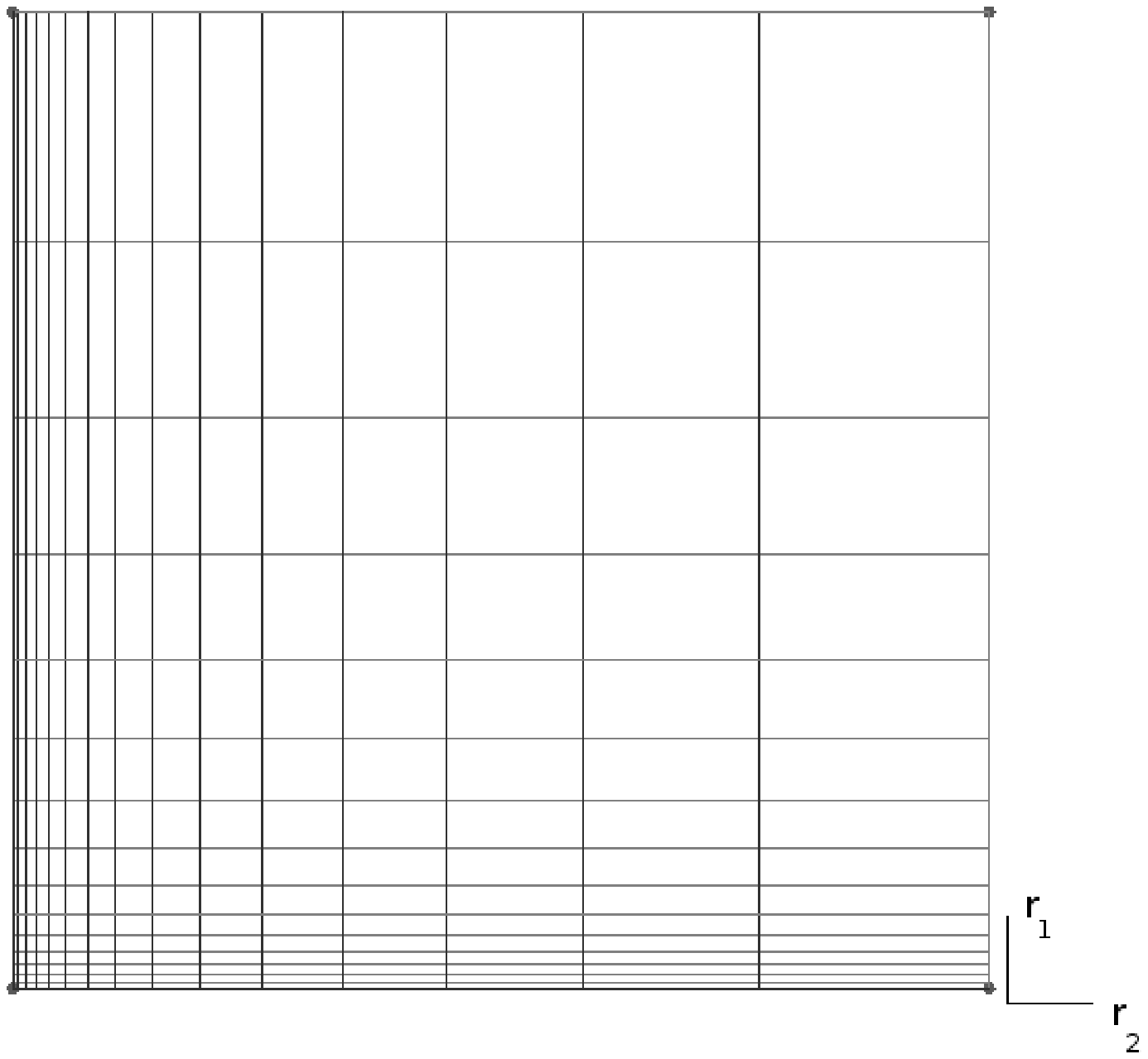} &
    \includegraphics[scale=.5]{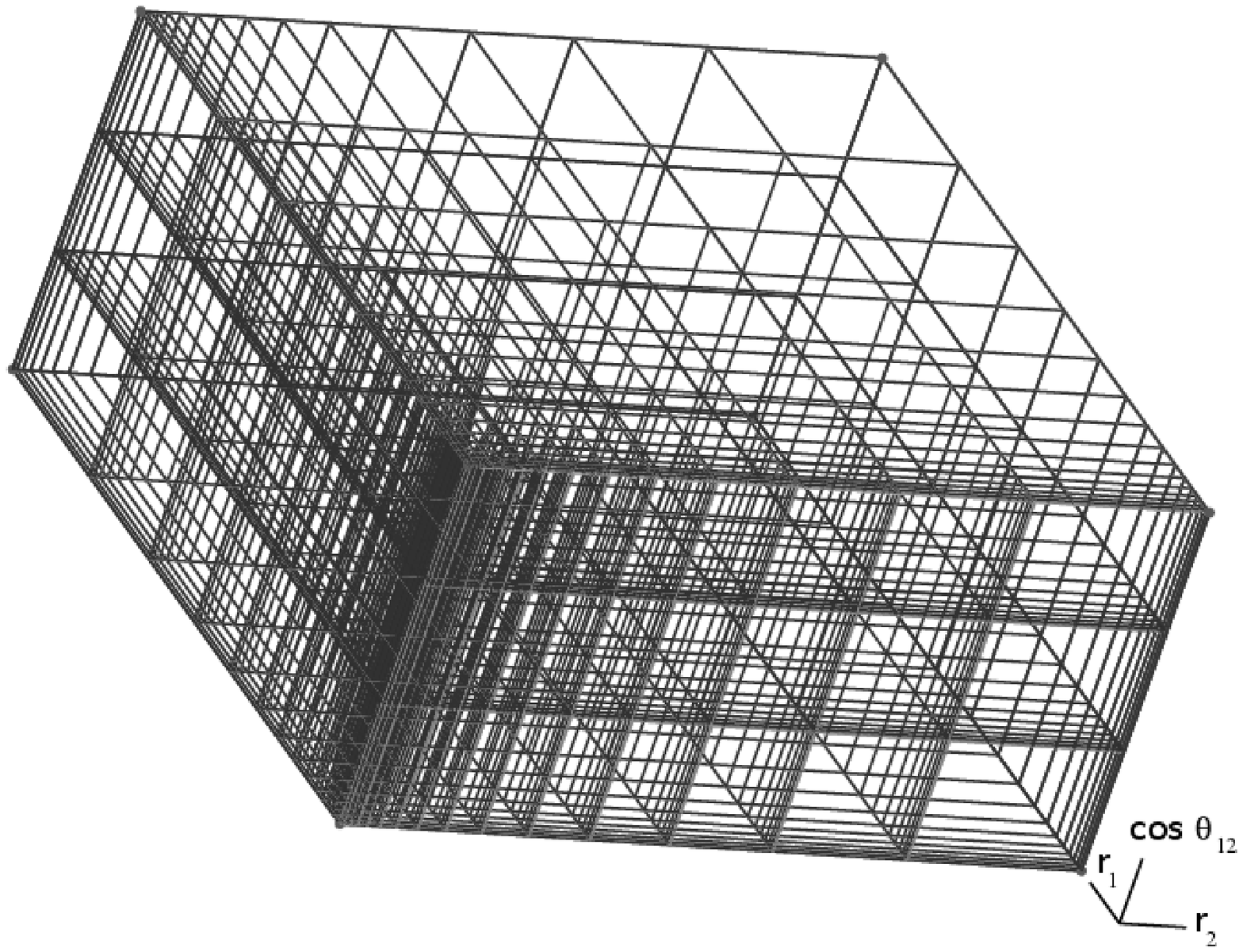} \\
  \end{tabular}
  \caption{example of an adaptive mesh \cite{GMSH} used in approximating the solution in three dimensions. In this case the number of elements is $N_{elements} = 15\times15\times3 = 675$, while the size of the element progressively increases by a factor of 1.3 as $r_{1,2}$ is increased}
  \label{fig:mesh}
\end{figure}
%%%%%%%%%%%%%%%%%%%%%%%%%%%%%%%%%%%%%%%%%%%%%%%%%

We can compare the one-particle formalism above with an exact formulation of the system. 
In 1930 Breit showed that the Schr\"{o}dinger equation for the two-electron atom becomes separable if one of the coordinate
axes is aligned with one of the radius vector \cite{Breit}. For the s state, the exact wavefunction depending on six total variables $\Psi(\vec{r_1},\vec{r_2})$ can be reduced to only three variables: the positions of the two electrons $r_1$ and $r_2$, and the angle between them $\theta_{12}$, viz
\begin{equation}\label{eqn:Exact}
  \int \int \int   \frac{1}{2}\frac{\partial \Psi^*}{\partial r_1} \frac{\partial \Psi}{\partial r_1} +  \frac{1}{2}\frac{\partial \Psi^*}{\partial r_2} \frac{\partial \Psi}{\partial r_2} +   \frac{1}{2}(\frac{1}{r_1^2}+\frac{1}{r_2^2})\frac{\partial \Psi}{\partial \cos\theta_{12}}\frac{\partial \Psi}{\partial \cos\theta_{12}} + \Psi^*(-\frac{Z}{r_1}-\frac{Z}{r_1}+\frac{1}{r_{12}}-E)\Psi  r_1^2 r_2^2 dr _1 dr_2 d\cos\theta_{12} = 0 
\end{equation}
where   $$r_{12} = \sqrt{r_1^2 + r_2^2 - r_1 r_2 \cos\theta_{12}}$$

We approximate the solution to the above Hamiltonian using FEM by discretizing each variable above independently as in \cite{Levin}. The boundary conditions implemented make the wavefunction vanish at a cutoff radius of $ r_{1,2} = 40$ a.u, and the angular range is between $ -1 \leq \cos \theta_{12} \leq 1$. Furthermore an adaptive grid was refined closer to the nuclei as shown in Fig.(\ref{fig:mesh}) to speed up calculations.  
In our implementation we use $C^1$-continuous basis, that is Hermite polynomials $N_{L \rightarrow R}(x) = 1-3x^2+2x^3 $ and $N_{L \leftarrow R}(x) =3x^2-2x^3$ to interpolate the left and right nodal values of the wavefunction and 
additionally $\bar{N}_{L \rightarrow R}(x) = h_x(x-2x^2+x^3)$ and $\bar{N}_{L \leftarrow R}(x)= h_x(x^3-x^2)$
to interpolate the derivatives of the wavefunction, where $h_x$ is the element width and $x = \{r_1,r_2,cos\theta_{12}\}$.
In this manner, the continuity of the wavefunction is enforced by using this basis. 
All the integrals involved in constructing the matrices were solved numerically using 10-point Gaussian quadratures, 
and in order to solve the generalized eigenvalue we used a sparse matrix solver in ARPACK++ \cite{ARPACK}. 

For the mesh shown in Fig.~\ref{fig:mesh}, we obtain a ground state energy of $E^{0}_{exact} = -2.7578$ using 
$C^0$-continuous basis and $E^{1}_{exact} = -2.8994$ using $C^1$-continuous basis.  
The latter value is close to the energy obtained by Levin using the same approach $E^{(He)} = -2.9032$ \cite{Levin}, which is very close given that we perform all integrals numerically.

\section{Results and Discussion}
In order to find the critical parameter we approximate the energy gap in the ground state numerically as
$$\Delta E_0(Z) \approx E_0^{2e}(Z) - E_0^{(e)}(Z),$$  where $E_0^{(2e)}$ is the binding energy obtained for two electrons
and $E_0^{(e)}$ is binding energy for a single electron. The theory of finite size scaling is applied to a helium-like Hamiltonian of the form,
\begin{eqnarray}
  \label{eqn:Hamiltonian}
  H & = & \underbrace{-\frac{1}{2}\nabla^2_1 -\frac{1}{2}\nabla^2_2 + \frac{1}{|r_1 -r_2|}}_{H_o} - 
  \underbrace{(\frac{Z}{r_1} + \frac{Z}{r_2})}_{V_Z},
\end{eqnarray}
 where $r_1$ and $r_2$ are the coordinates of the two electrons referenced from the nuclei, and Z is the 
charge of the nucleus, this will play the role of the critical parameter. For reference, we note that the 
transformation $r\rightarrow r/Z$ takes Eq. (\ref{eqn:Hamiltonian}) into:

\begin{eqnarray}
  \label{eqn:Hamiltonian_trans}
  H(r\rightarrow r/Z) & = & Z^2[\underbrace{-\frac{1}{2}\nabla^2_1 -\frac{1}{2}\nabla^2_2 - (\frac{1}{r_1} + \frac{1}{r_2})}_{H'_o} + \underbrace{\frac{1}{Z} \frac{1}{|r_1 -r_2|}}_{V'_Z}]
\end{eqnarray}
this form is often used in the literature to obtain the critical parameter $\lambda_c =1/Z_c$ (see \cite{Stillinger,Baker}). 
Below we outline a general procedure that can be used to combine FSS with numerical methods that expand a given 
wavefunction in the Hilbert space in any complete basis at a truncated level N.
%%%%%%%%%%%%%%%%%%%%%
\par
\begin{figure}
  \centering
  \begin{tabular}{cc}
    \par\hspace{-7mm} \includegraphics[scale=.77]{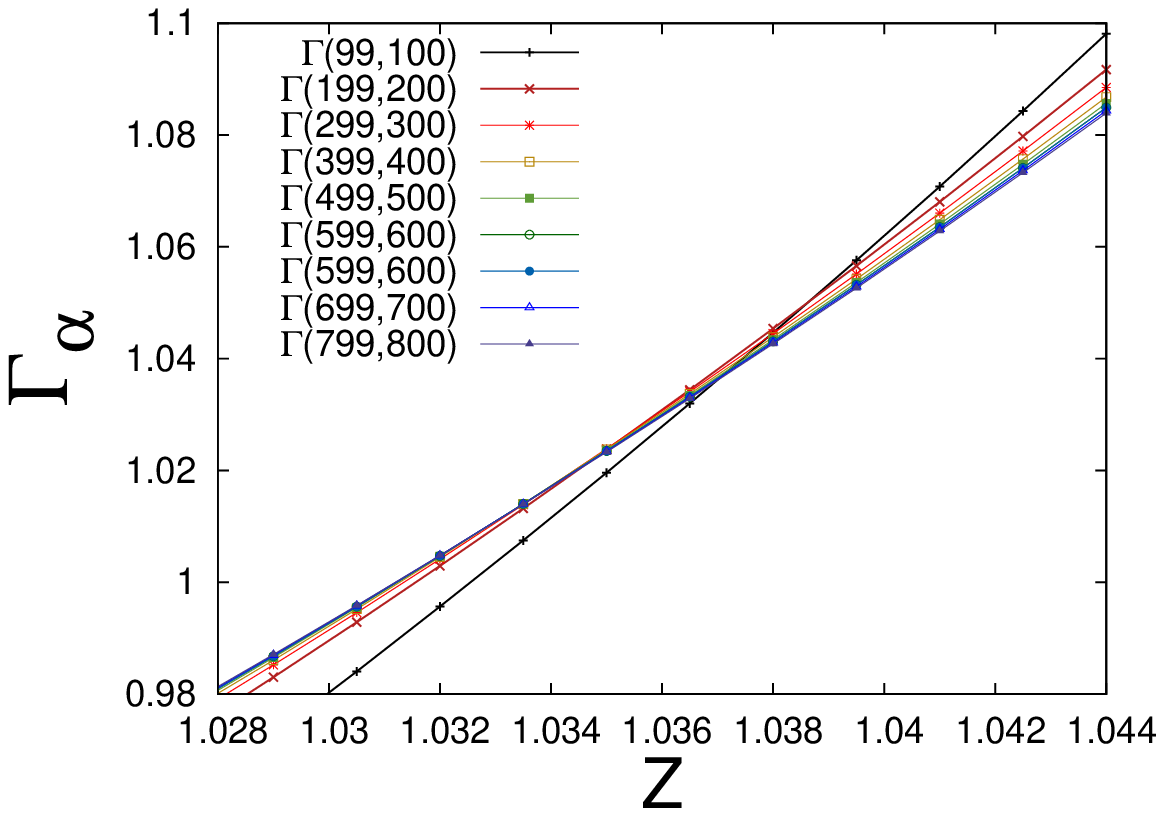} &
    \par\hspace{-7mm} \includegraphics[scale=.77]{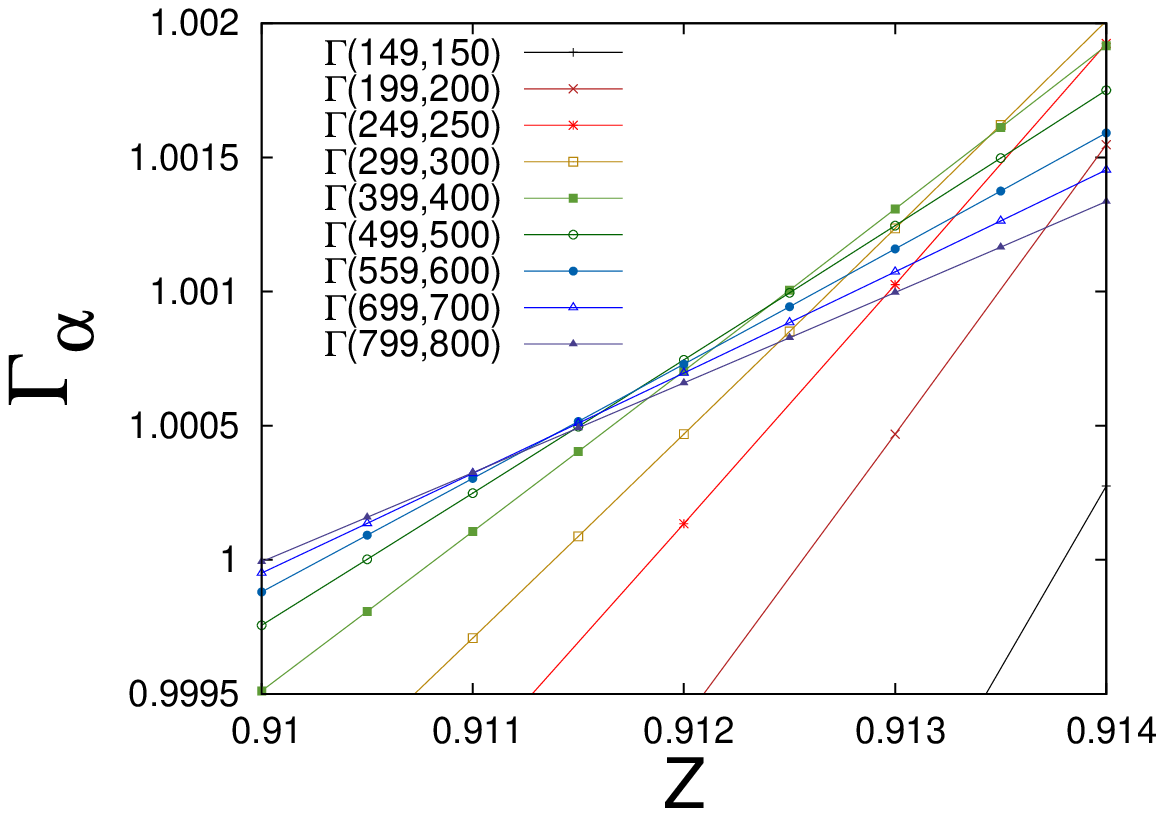} \\
    \par\hspace{3mm} (a)  &   \par\hspace{3 mm} (b)  \\
    \par\hspace{-7mm} \includegraphics[scale=.77]{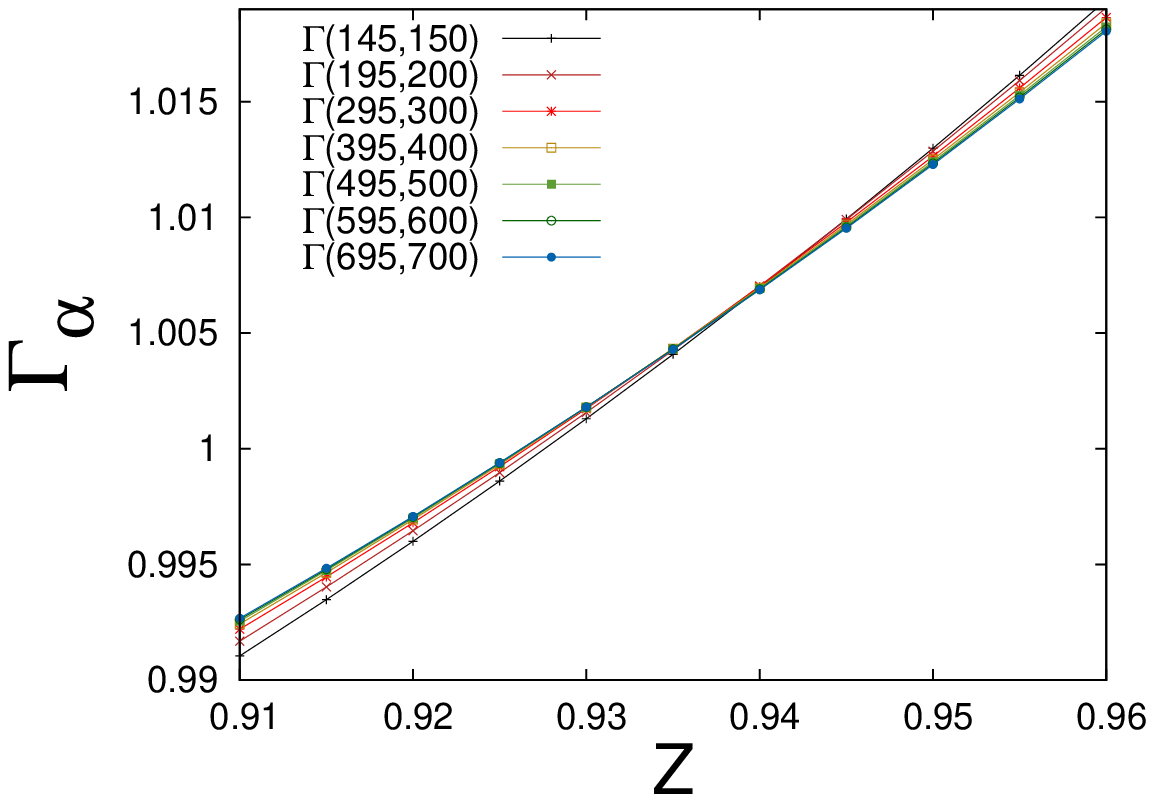} &
    \par\hspace{-7mm} \includegraphics[scale=.77]{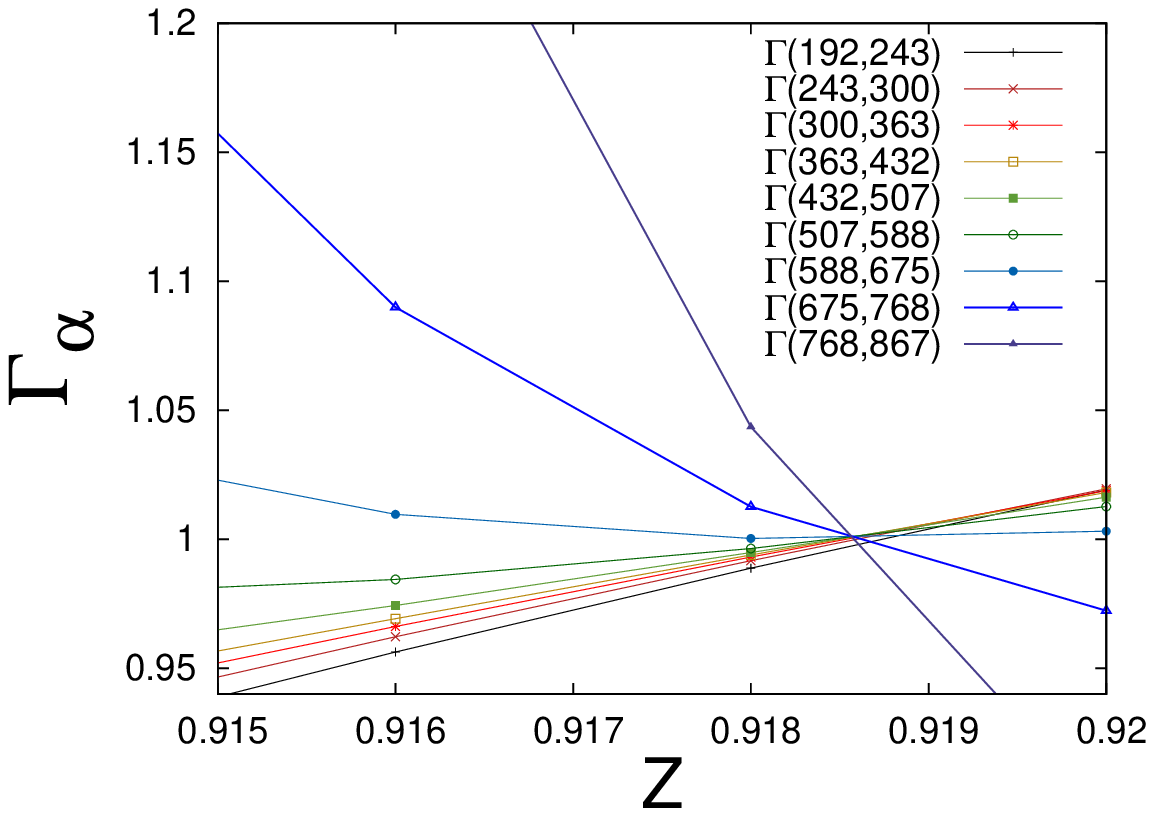} \\
    \par\hspace{3mm} (c)  &   \par\hspace{3 mm} (d)  \\
  \end{tabular}
  \caption{(Color online) Plot of $\Gamma_\alpha$ (Eq. ~\ref{eq:gammafunc}) obtained by using FSS method as a function of the charge Z for increasing number of elements (N) for: (a) Hartree-Fock only, (b) Total Energy (HF + correlation energy), (c) LDA approximation,  and (d) exact formulation}
  \label{fig:Xssing}
\end{figure}

\begin{figure}[htpb]
%\par\vspace{-10mm}
  \centering
  \includegraphics[scale=.90]{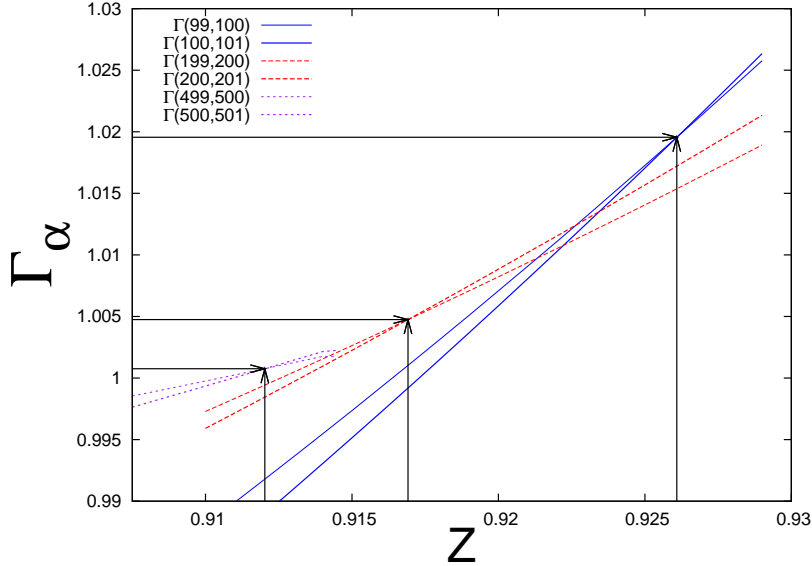}
  \caption{The crossing of $\Gamma_\alpha(Z,N-1,N)$ and $\Gamma_\alpha(Z,N,N+1)$ as N is increased using Total Energy approximation in HF. The crossing has been slightly enhanced in order to demonstrate the dependence of N.}
  \label{fig:xssing}
\end{figure}

\begin{figure}
\par\vspace{5mm}
  \centering
  \begin{tabular}{ccc}
    \includegraphics[scale=.485]{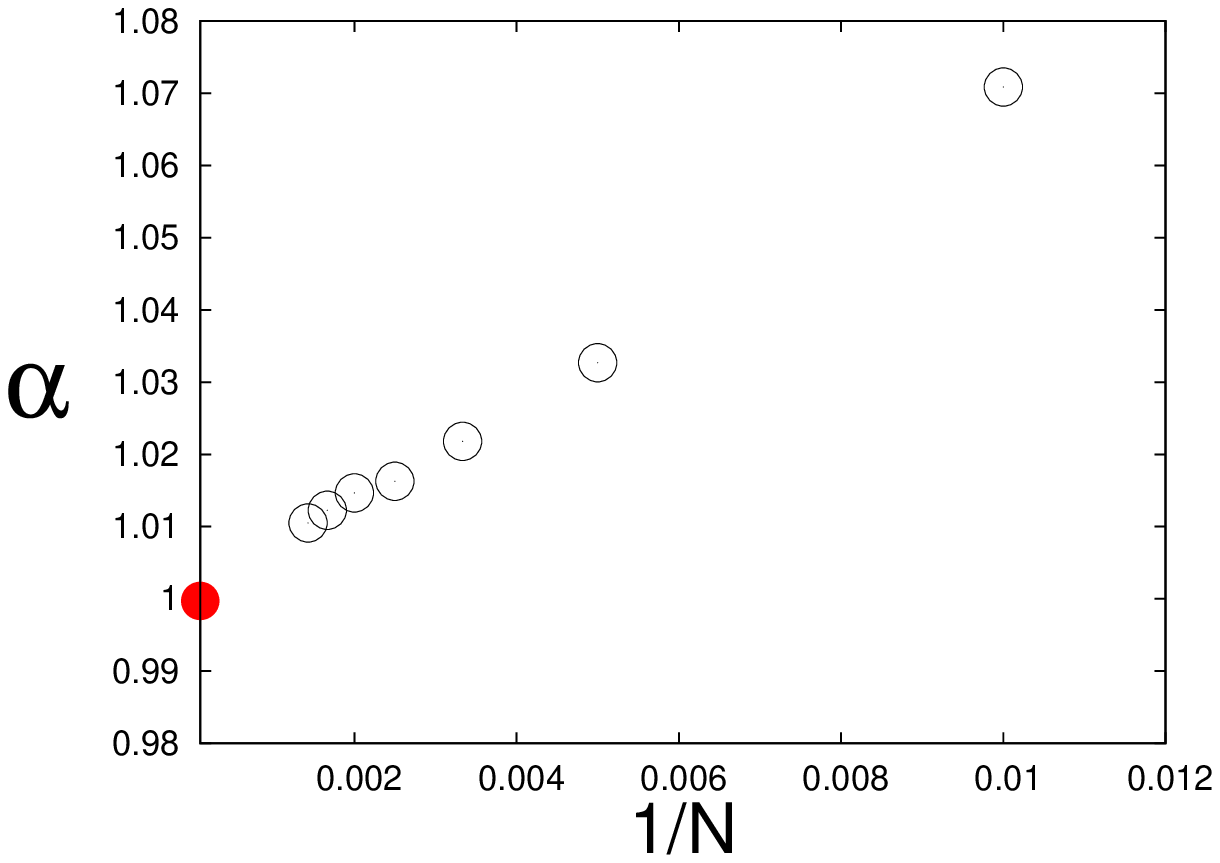} &
    \includegraphics[scale=.485]{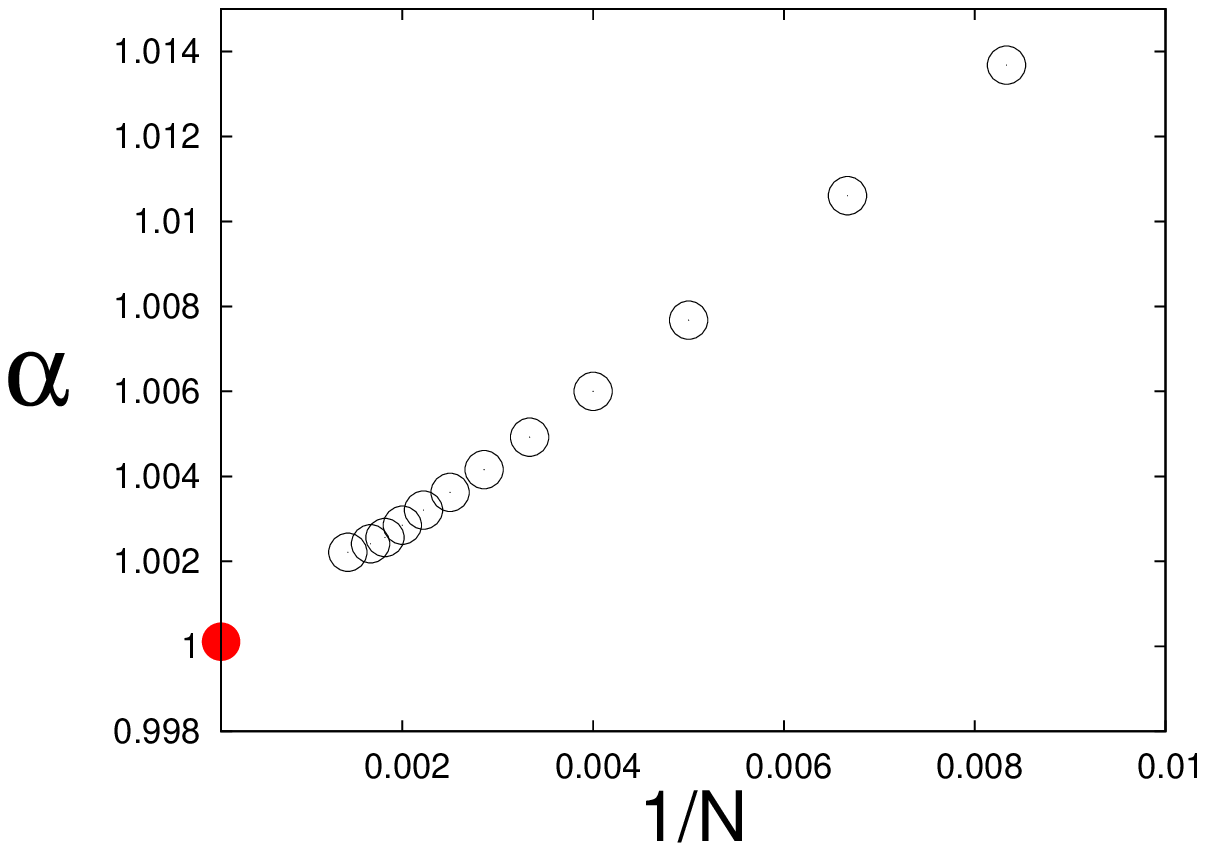} &
    \includegraphics[scale=.485]{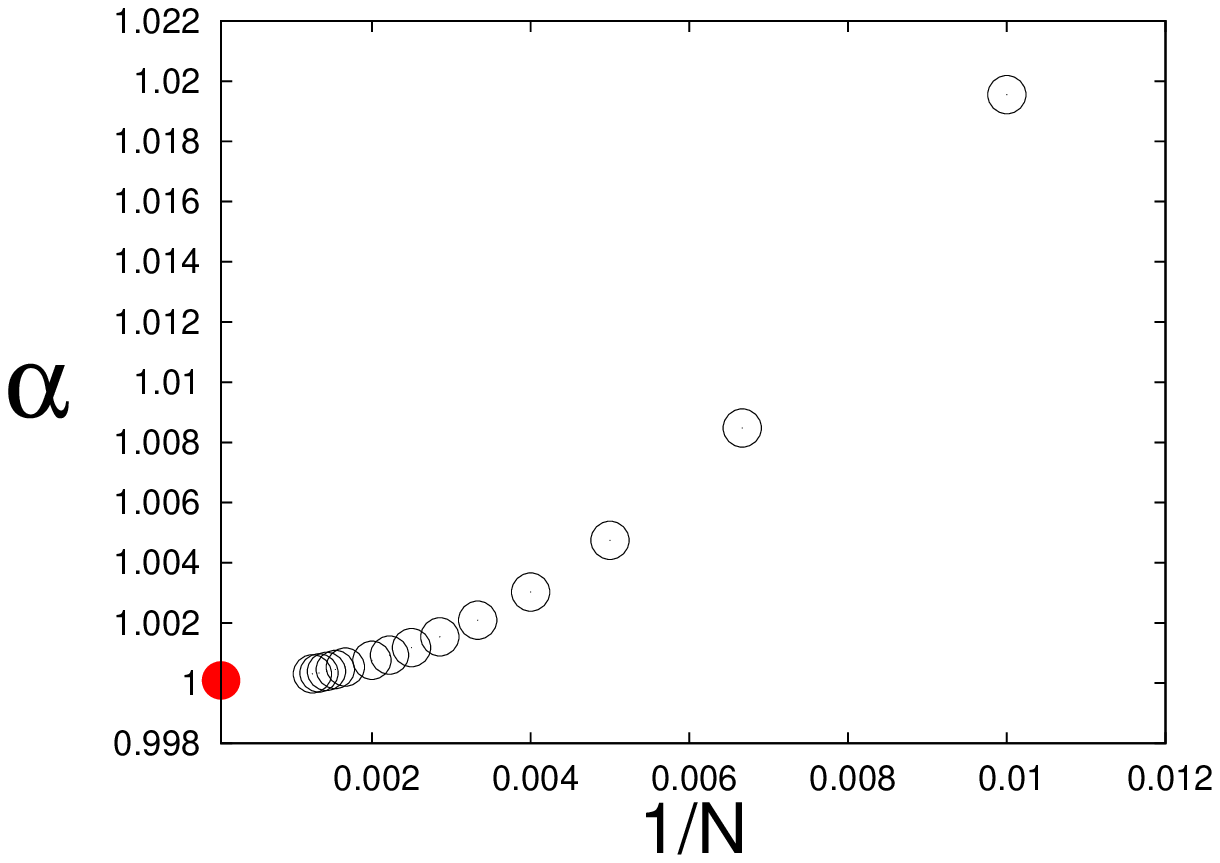} \\
     & \\
     & \\
    \includegraphics[scale=.485]{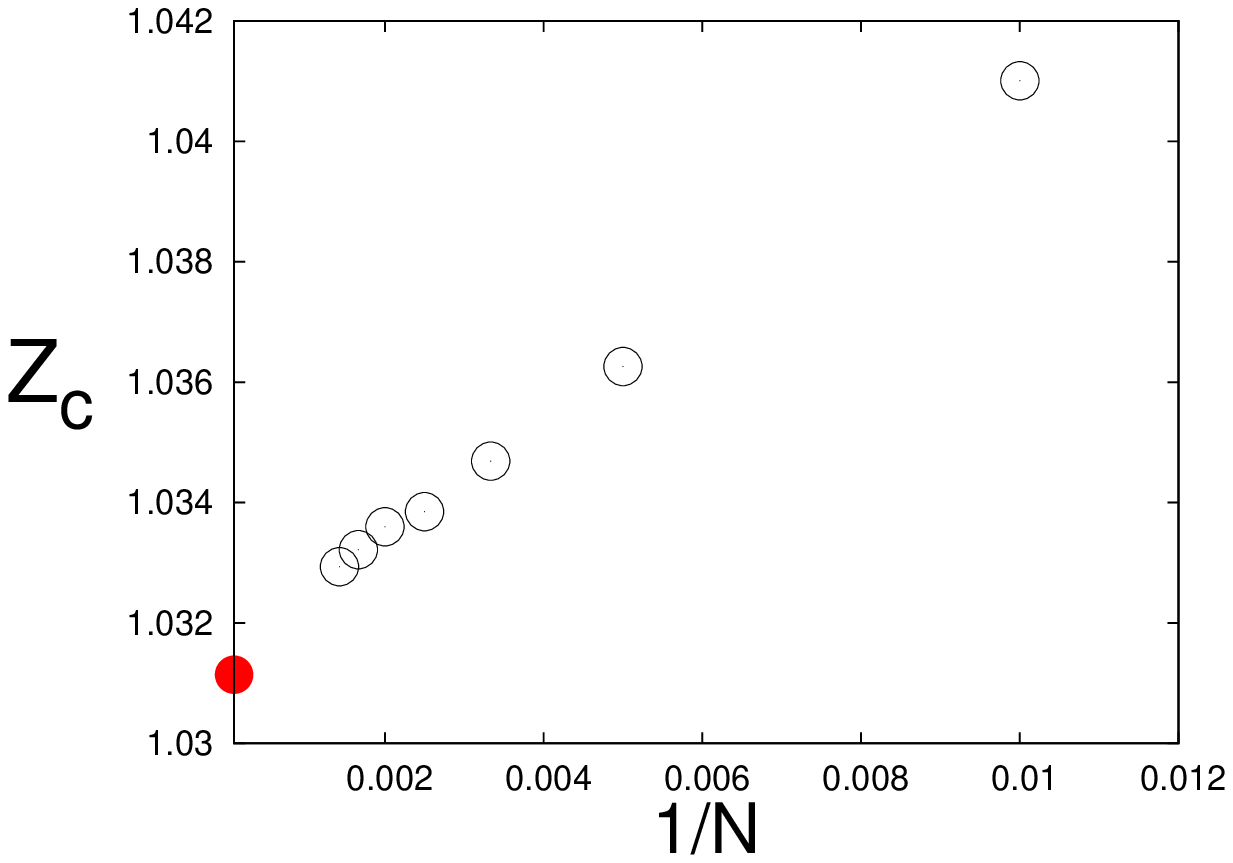} &
    \includegraphics[scale=.485]{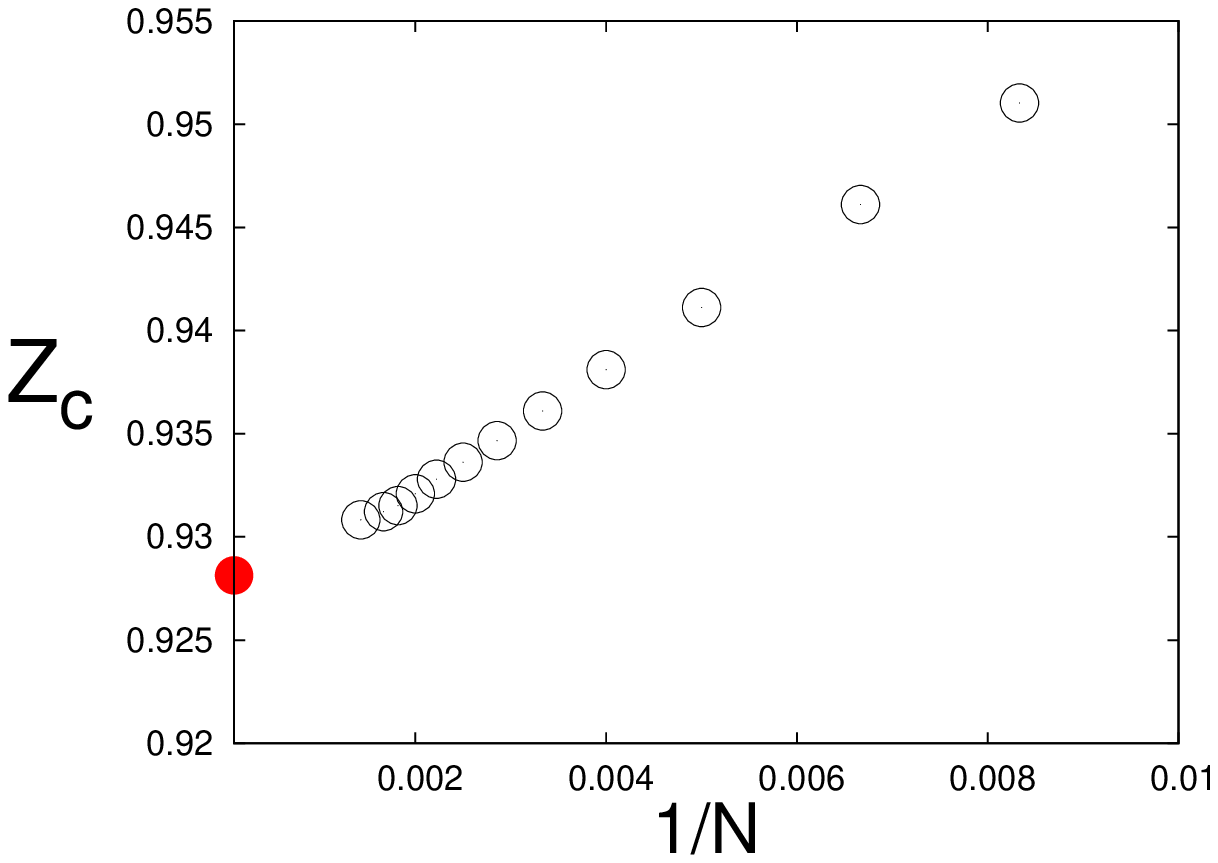} &
    \includegraphics[scale=.485]{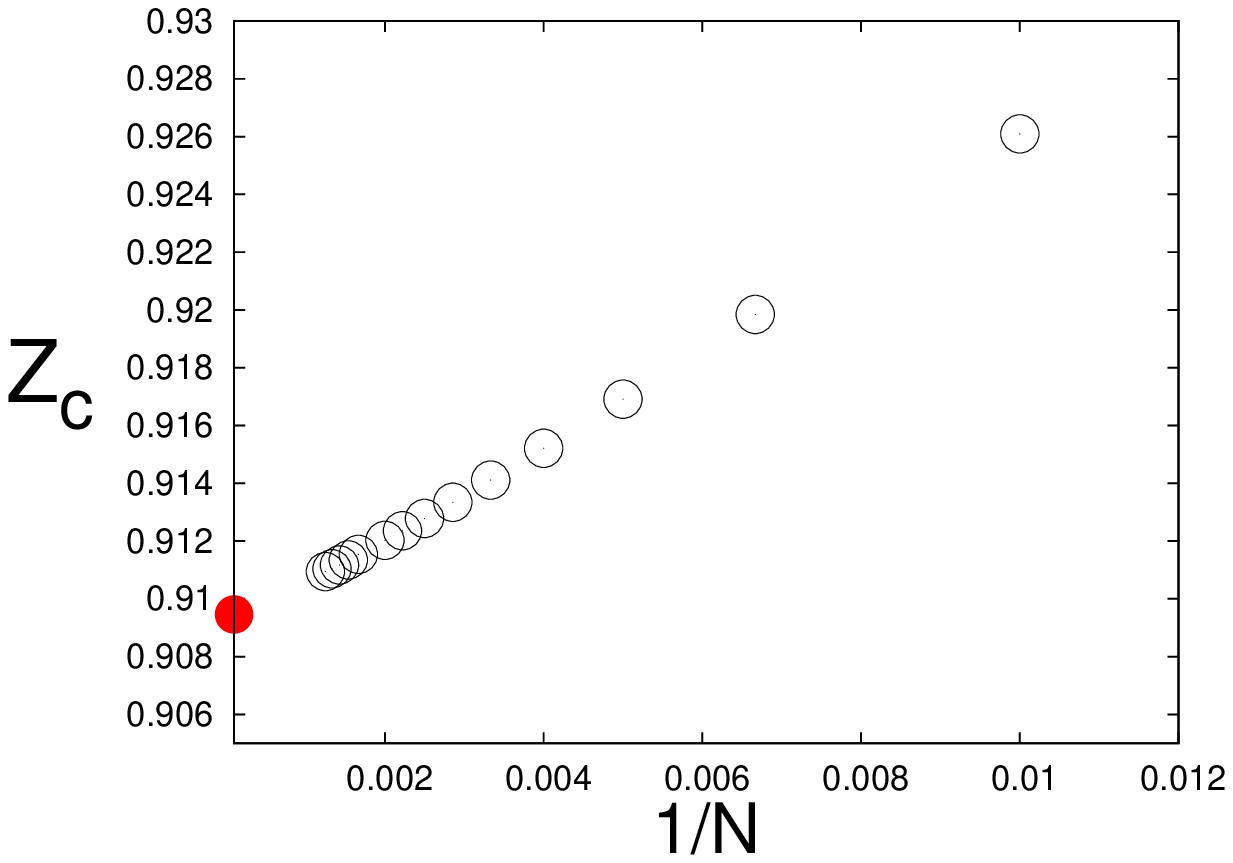} \\
    \par\hspace{10mm} (a)  &   \par\hspace{10mm} (b)   &   \par\hspace{10mm} (c) \\  
\\
  \end{tabular}
  \caption{Extrapolated values for the critical exponent $\alpha$ (top) and $Z_c$ (bottom) for: (a) HF approximation, (b) LDA approximation, and (c) Total energy.  The filled dots are the extrapolated values for the $ 1/N \rightarrow 0$ limit}
  \label{fig:extrapolation_meanfield}
\end{figure}

\begin{figure}
  \centering
  \begin{tabular}{cc}
    \includegraphics[scale=.565]{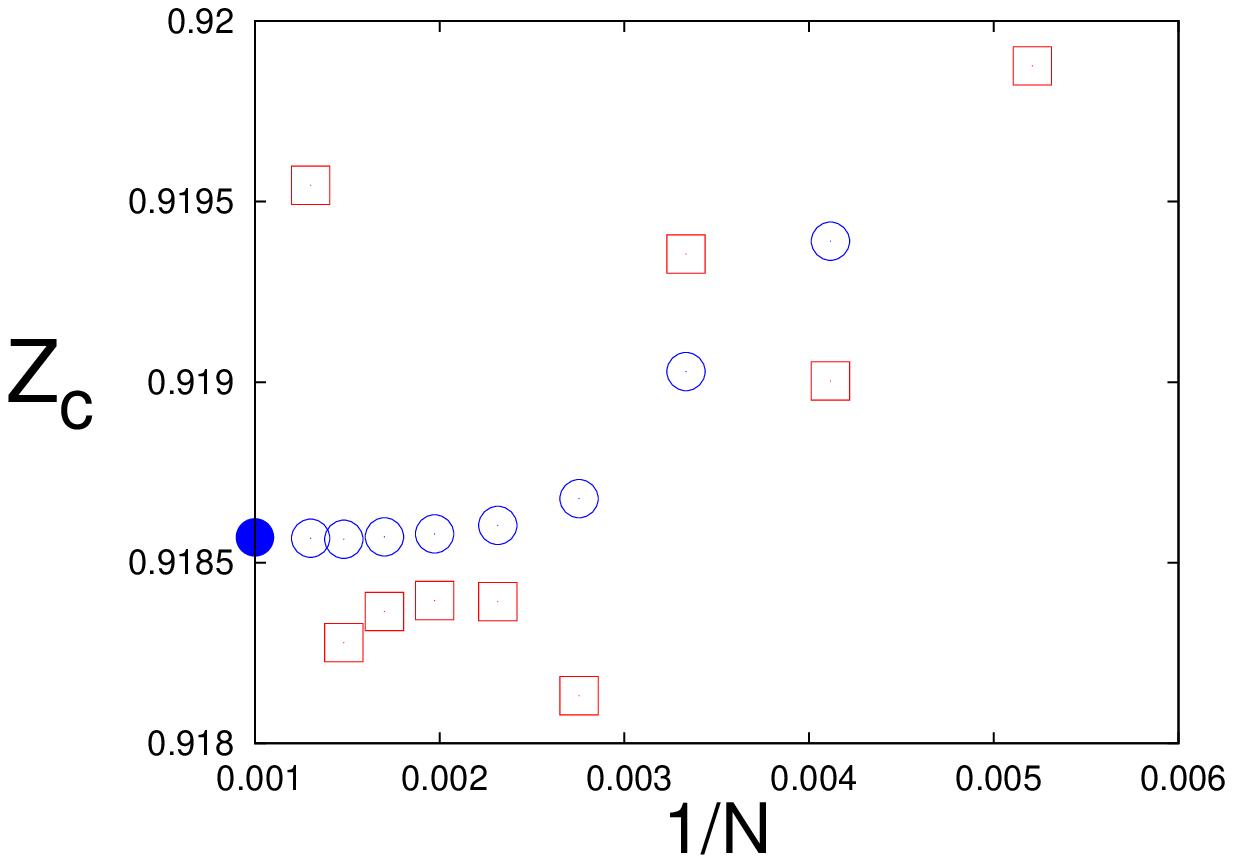} &
    \includegraphics[scale=.565]{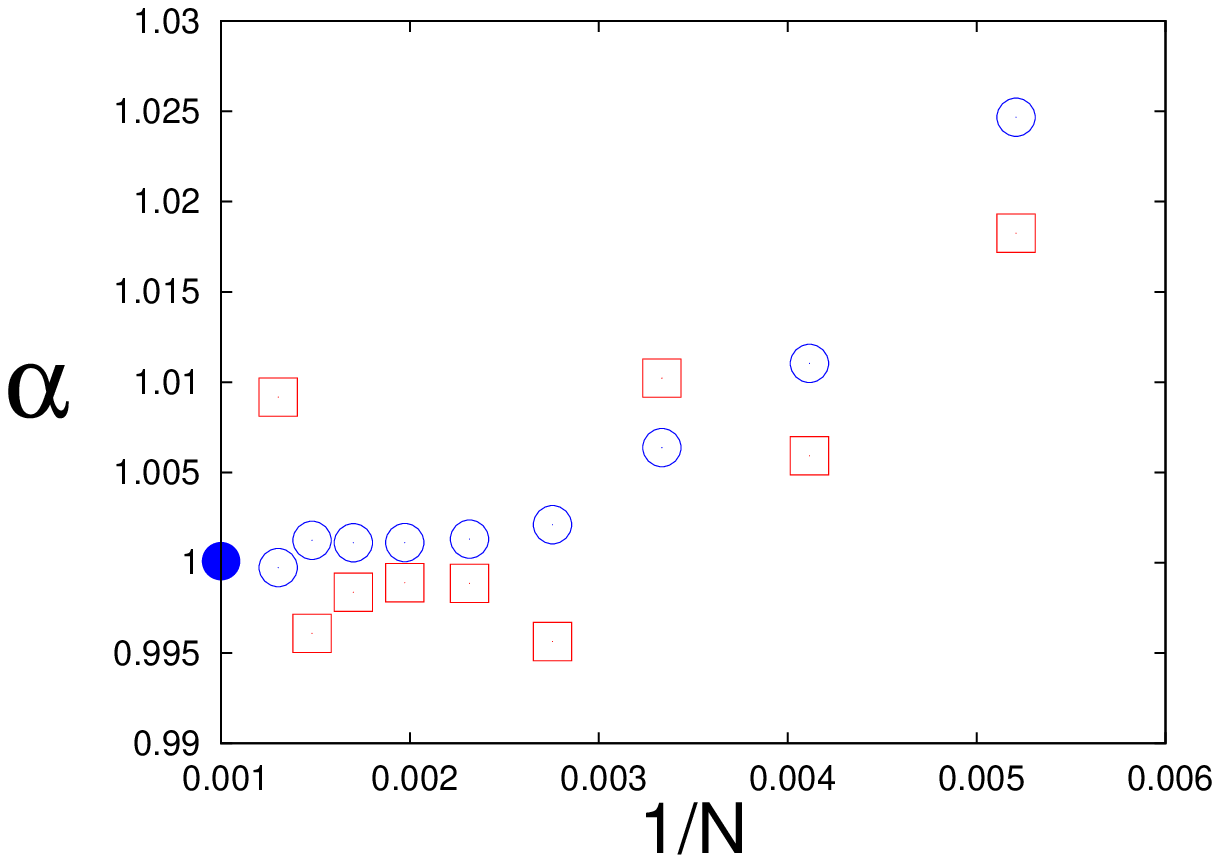} \\
  \end{tabular}
  \caption{Extrapolated values for the critical exponent $Z_c$ and $\alpha$ in the $ 1/N \rightarrow 0$ limit solving the system exactly in 3D. 
    We construct the $\Gamma_\alpha$ function using \emph{Hermite} interpolation polynomials for two potentials: 
    $V_Z = -Z(\frac{1}{r_1} + \frac{1}{r_2}) $ (denoted by $\square$) and 
    $V_Z = \frac{1/Z}{\sqrt{r_1^2 + r_2^2 - r_1 r_2 \cos\theta_{12}}}$ (denoted by $\bigcirc$); 
    the latter is obtained by making the transformation $r\rightarrow r/Z$ in the Hamiltonian.
    We checked that the fluctuations seen in the figure vanish by replacing the numerical approximation to the binding energy with the analytical form, 
    $E^{(e)} = Z^2/2$, while the extrapolated value remains unchanged.}
  \label{fig:extrapolation_exact}
\end{figure}

\vspace{5mm}
\underline{\textsc{General Procedure}}: \\
\begin{enumerate}
\item Compute the expectation value of $\left<H\right>^{(N)}$ and $\left< \frac{dH_Z}{dZ} \right>^{(N)}$ for various values of N (number of elements). In this case the Hellmann-Feynman theorem (see Eq.~\ref{eqn:Hamiltonian}) becomes useful since in this case $H_Z$ is linear in $Z$ and the derivative is quite simple  $\left< \frac{dH_Z}{dZ} \right>^{(N)} \rightarrow \left< \frac{dV_Z}{dZ} \right>^{(N)}$.
\item Construct the gamma function $\Gamma_\alpha(Z,N,N')$ (see Eq. \ref{eq:gammafunc}) for several values of $(N, N')$ vs $Z$.
At criticality, the gamma function $\Gamma_\alpha(Z,N,N')$ becomes independent
of N and N' and the curves will cross. In Fig.\ref{fig:Xssing} we show the gamma function we
obtain by using different number of elements and how it begins to cross for various approximations to the energy.
\item In order to go to the complete basis set limit  $\left<H\right>^{\infty}$, 
we have to let $(N,N') \rightarrow \infty$. The best choice is to keep
$N-N'$ minimal \cite{nightingale1,privman}. Then, we proceed as follows: We solve $\left<H\right>^{(N)}$ for (N-1, N , N+1). 
This provides us with two gamma curves: $\Gamma_\alpha(Z,N-1,N)$ and $\Gamma_\alpha(Z,N,N+1)$, 
with minimum $N-N'$. The Crossing of these two curves gives us the pseudo-critical parameters $\alpha^{(N)}$ and $Z^{(N)}_c$
with a truncated basis set number N. Fig.~\ref{fig:xssing} illustrates the N dependence of the crossing.
\item In order to extrapolate the values $\alpha^{(N)}$ and $Z_c^{(N)}$ to the infinite N limit, we use
the algorithm of Burlish and Stoer \cite{BS}. Fig.s~\ref{fig:extrapolation_meanfield} and \ref{fig:extrapolation_exact} show their behavior by systematically increasing N for the mean-field approximations and exact formulation, respectively.\\
\end{enumerate}

%%%%%%%%%%%%%%%%%%%%%%

%%%%%%%%%%%%%%%%%%%%%%

While finding values $\alpha^{(N)}$ and $Z^{(N)}$ and extrapolating to the $ 1/N \rightarrow 0$ limit 
some numerical fluctuations can arise depending on the approximation used. 
We discuss some ways of dealing with these fluctuations. First, 
one can improve the basis used from $C^0$-continuous polynomials to $C^1$-continuous polynomials
in order to obtain greater accuracy in the calculations, however, matrices double in size increasing 
the memory cost as well as the computing time. Second, one can 
construct a gamma function from a potential in the Hamiltonian that explicitly takes into 
account all variables. For example in the three dimensional (3D) case we found that it is better to use 
the transformed Hamiltonian (Eq.\ref{eqn:Hamiltonian_trans}) where the perturbation $V_\lambda$ is given 
by $V_{1/Z} = \frac{1/Z}{\sqrt{r_1^2 + r_2^2 - r_1 r_2 \cos\theta_{12}}}$ 
rather than $V_Z = \frac{Z}{r_1} + \frac{Z}{r_2} $. This does not change the 
physics of the system since both expressions can be used to find the
critical value: either $1/Z_c$ or $Z_c$, which corresponds to one of the
electrons going from a bound state into the continuum. The first perturbation $V_{1/Z}$
takes into account all three variables when the expectation value is computed whereas the latter $V_{Z}$
introduces numerical errors in the expectation value. 

%%%%%%%%%%%%%%%%%%%%%%%%%%%%%%%%%%%%%%%%%%%%%%%%%%%%%%

%%%%%%%%%%%%%%%%%%%%%%

\begin{table}[htp]\bf
  \caption{Critical parameters extrapolated from Fig.(\ref{fig:extrapolation_meanfield}) and  Fig.(\ref{fig:extrapolation_exact})}
  \centering
  \begin{tabular}{|c  c  c  c  c |  c  | }
    \hline\hline
    $    $    &HF   &~~LDA  &Total Energy & Exact & Reference \cite{Baker,PSK}\\
    \hline
    \underline{\small{critical charge}}  & & & & &\\
    $Z_c$ &1.03114 &~~0.92808 & 0.90946 & 0.91857~ & 0.91103\\
    \underline{\small{critical exponent}}  & & & & &\\
    $\alpha$  &0.99971 &~~1.00011  &1.00009 & 1.00082~ & 1.00000\\
    \hline
    \hline
  \end{tabular}
  \label{table:results}
\end{table}

%%%%%%%%%%%%%%%%%%%%%%%%%%%%%%%%%%%%%%%%%%%%%%%%

%%%%%%%%%%%%%%%%%%%%%%%%%%%%%%%%%%%%%%%%%%%%%%%%
%%%%%%%%%%%%%%%%%%%%%%%%%%%%%%%%%%%%%%%%%%%%%%%%%%%%

Last, another method is to increase the minimum difference $|N-N'| \geq \delta$ used in finding a crossing between $\Gamma_\alpha(Z,N-\delta,N)$  and $\Gamma_\alpha(Z,N,N+\delta)$ until a systematic crossing is obtained for all values of N; 
 in the LDA approximations we used $|N-N'| = 5$, since the Gamma curves change very little as the number of elements increases (see Fig.\ref{fig:Xssing}-c), 
whereas in the 3D formulation we used $ |\sqrt{N/3} -\sqrt{N'/3} | =1 $ since in this case the number of elements increases as $N_{elements} = N_{r_1}\times N_{r_2} \times N_{\cos\theta}$, and $N_{\cos\theta}$ was kept fixed at three. 
In solving the self-consistent equations [ Eqs.(\ref{eqn:HF}) and (\ref{eqn:KS})], we found that the region below $Z < 1$ 
became problematic when updating the self-consistent equations. This is because as the nuclear charge decreases, the
wavefunction decay becomes slower with respect to r and the cutoff radius might lead 
to unphysical solutions (i.e. solutions which depend on $r_{cut}$). 
To get around this problem, we found it is useful to perturb the wavefunction with the arithmetic 
average of the current and predicted wavefunctions, rather than just the predicted wavefunction. 
In this manner, we can avoid converging to a local minima that corresponds to an unphysical solution. 

In phase transition theory it is well known that using mean field theory (for dimensionality $\leq 4$) does not give correct values near a critical point. This is indeed the case by using only the Hartree-Fock approximation (see Table \ref{table:results} ), however we see 
that incorporating the exchange-correlation energy within the LDA approximation improves substantially the critical charge within 2\% 
of the accepted value and within 1\% using Total Energy approximation and exact formulation. It is interesting to note that all approaches give a consistent estimate for the critical exponent, regardless of the approximation used. 
This study demonstrates that Finite-element method is a viable tool that can be applied in combination with
finite size scaling in order to obtain good estimates of critical parameters starting 
with ab initio calculations for electronic calculations such as DFT and HF and compare well 
with other methods that are deemed more precise. 

We now check the consistency of our approach. 
In the theory of continuous phase transitions, a quantity diverging in the thermodynamic limit 
($N \rightarrow \infty$), becomes a regular function for finite N, viz  $\left<O\right>^{(N)} \sim N^{-\mu_o/\nu}$.
We can check that this assumption is indeed satisfied for any value N.
In the infinite basis limit we have the following form \emph{near criticality}: $\left<O\right>^{(\infty)} \sim  (Z - Z_c)^{\mu_o}$ (Eq.\ref{eqn:gap}).
In this manner, can obtain the $m^{th}$ derivative of the energy near criticality given that we have a limiting
expression for the energy, viz
\begin{eqnarray}
  E_Z & \sim & (Z - Z_c)^\alpha  \Rightarrow \frac{d^m E_Z}{dZ^m}  \sim (Z - Z_c)^{\alpha-m}
\end{eqnarray}
In a finite basis the above expression becomes a power law in the number of basis with the following characteristic exponents.
\begin{equation}
  E^{(N)}_Z \sim N^{-\alpha/\nu}~~~and~~~\frac{d^m E^{(N)}_Z}{dZ^m} \sim N^{-(\alpha-m)/\nu}
\end{equation}

%\newpage
Excluding the trivial case N=0, one can therefore use Taylor expansion around $|Z-Z_c|\leq 1$.
\begin{eqnarray}\label{eq:dataColl}
  E^{(N)}_Z & = & \sum_m \frac{d^mE^{(N)}_Z}{dZ^m} \cdot \frac{(Z - Z_c)^m}{m!} \\
  & = & \sum_m N^{(-\alpha+m)/\nu} \frac{(Z - Z_c)^m}{m!} \\
  & = &N^{-\alpha/\nu} \underbrace{\sum_m \frac{[N^{1/\nu}(Z - Z_c)]^m}{m!}} \\
  & = &N^{-\alpha/\nu} G_0(N^{1/\nu}(Z - Z_c))
\end{eqnarray}

\begin{figure}[htpb]
  \centering
  \begin{tabular}{cc}
      \includegraphics[scale=0.665]{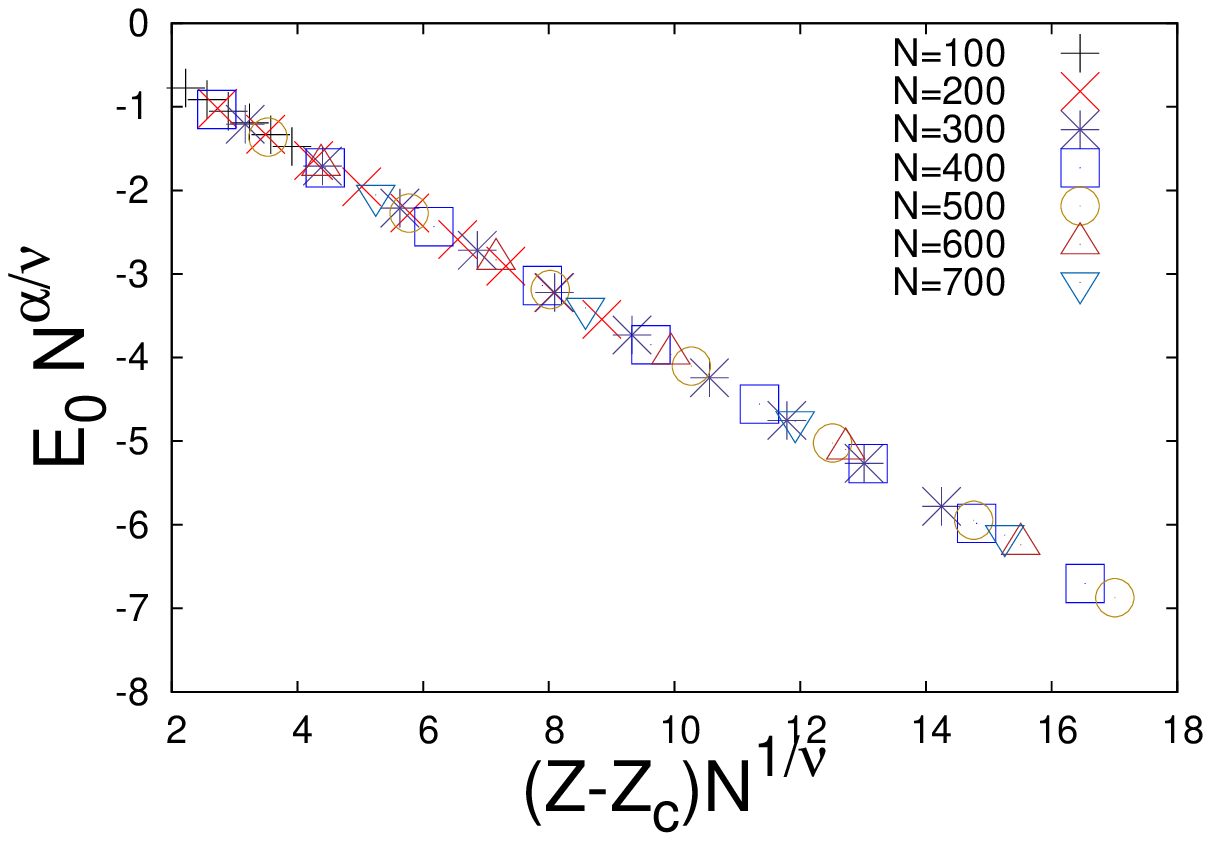}&
      \includegraphics[scale=0.665]{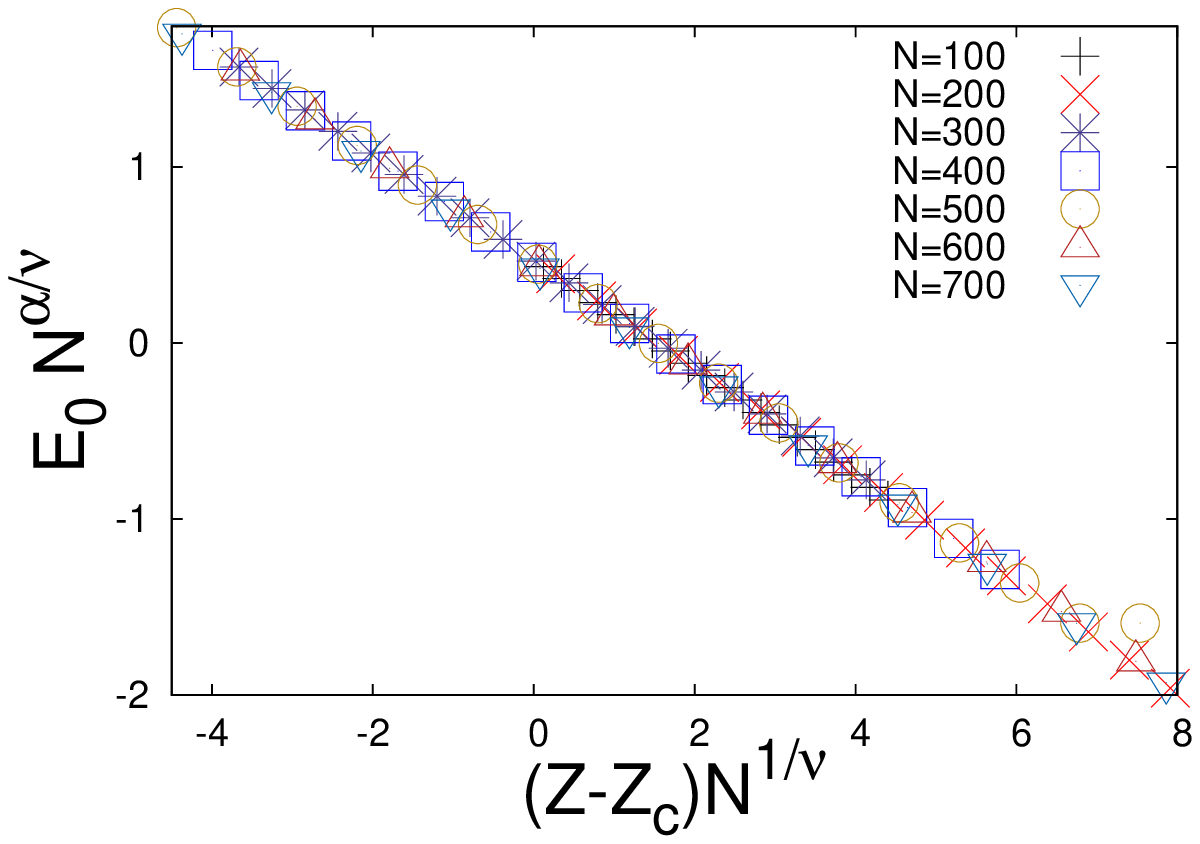} \\
      \par\hspace{10mm} (a)  &   \par\hspace{10mm} (b)  \\
      \includegraphics[scale=0.665]{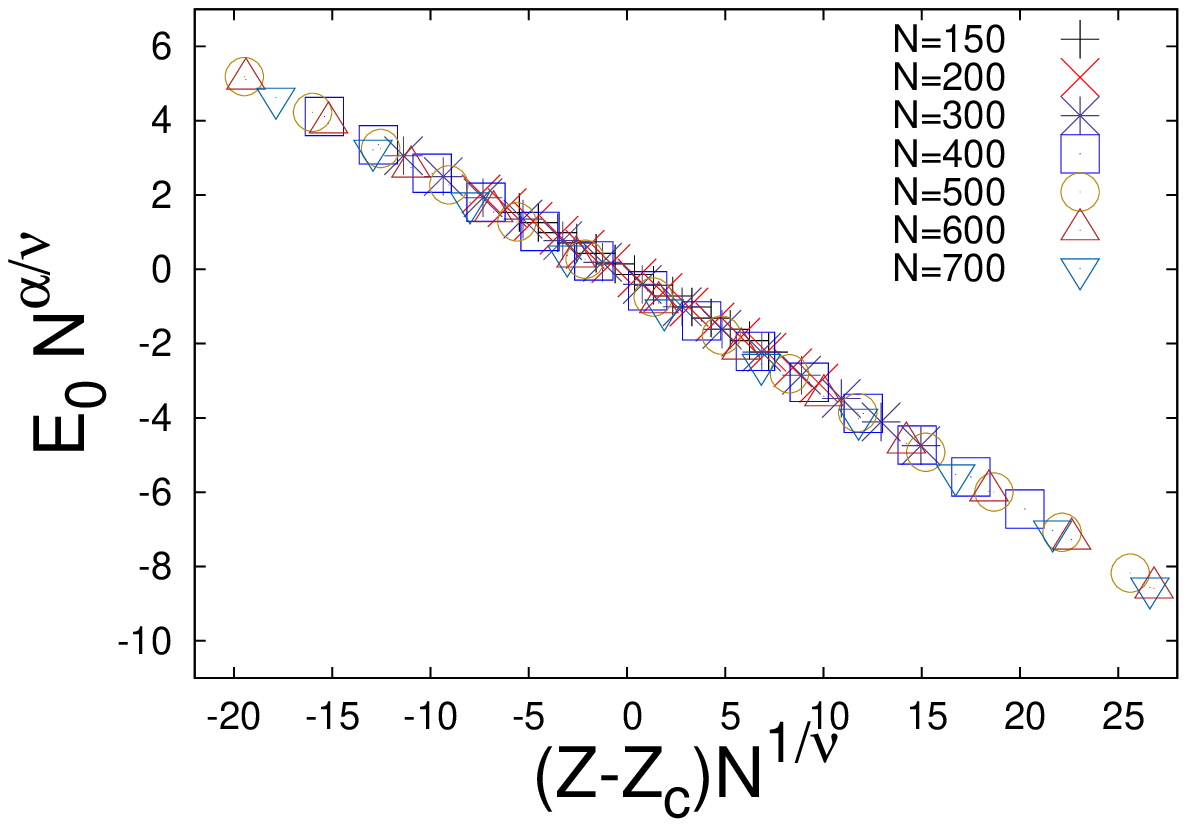}&
      \includegraphics[scale=0.665]{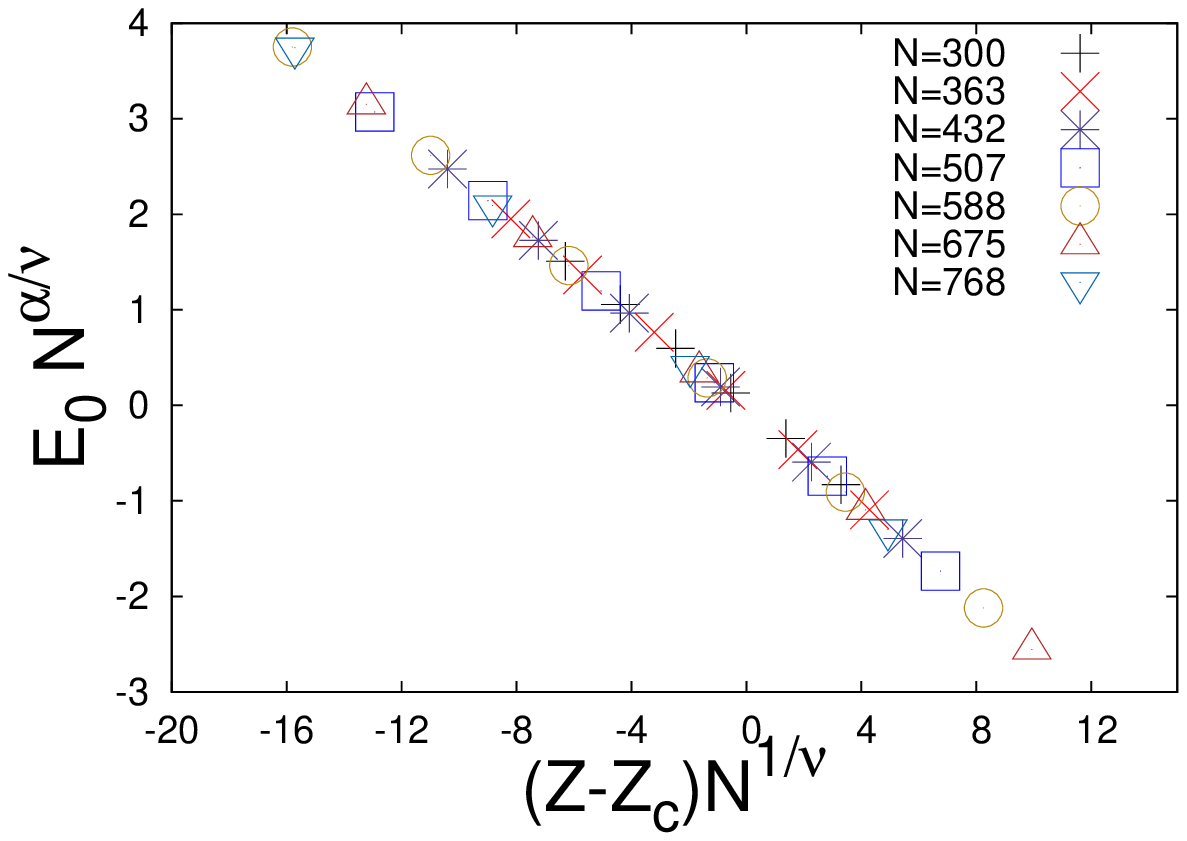}\\
      \par\hspace{10mm} (c)  &   \par\hspace{10mm} (d)  \\
  \end{tabular}
  \caption{(Color online) Data collapse for various values of N using: (a) HF, (b) Total Energy, (c) LDA approximation, and (d) exact formulation. The value from this collapse study that give us the best correlation length exponent $\nu \approx .85 \pm .05$, it should be emphasized that the values $\alpha$ and $Z_c$ used in making these figures are the corresponding values for each approximation (i.e. those shown in Table ~\ref{table:results})}
  \label{fig:figure4}
\end{figure}

In this manner we obtain a universal scaling function $G_0$. 
Plotting $ E^{(N)} N^{\alpha/\nu} $ vs the argument of the function $G_0$ would confirm whether the 
scaling function depends on N or not. Fig.~\ref{fig:figure4} shows that all quantities obtained from different N values do in fact 
collapse into one curve, and this confirms the scaling hypothesis for FSS with FEM (Eq.\ref{eq:fss_QM}).
In order to make the data collapse, it should be emphasized that the values $\alpha$ and $Z_c$ used 
to produce Fig.\ref{fig:figure4} correspond to those obtained in Table~\ref{table:results}.
If other values are used, the curves will \emph{not collapse} and they will be distributed in a family of curves (N dependent)
all deviating away from the unique curve shown in Fig.\ref{fig:figure4}. The correlation length exponent was varied until the best collapse curve was found. This occurred around $\nu \approx .85 \pm .05$ for all implementations.

%%%%%%%%%%%%%%%%%%%%%%%%%%%%%%%%%%%%%%%%%%%%%%
%%%%%%%%%%%%%%%%%%%%%%%%%%%%%%%%%%%%%%%%%%%%%%

The solution to a set of Hartree-Fock or Kohn-Sham equations for large number of atoms becomes the limiting step,
and it is here that the FEM can extend the range of sizes that can be
investigated by other current means. The efficiency of FEM in the context of 
large-scale calculations derives from the strict locality of the FEM basis in real space, 
thus minimizing the need for extensive interprocessor communications on massively parallel 
computational architectures \cite{pask,pask2}. The work of Pask {\it et al} \cite{pask,pask2} has formulated a real-space DFT approach 
for condensed matter applications without the need of using expensive Fourier transforms.
%The idea is that the ionic potential is split into a local part $V_I^{(loc)}$ and a non-local part $V_I^{(nl)}$,
%thereby the calculation of the long range potential is replaced with a short range charge density that
%generates it and the long-range effects are incorporated into the {\it boundary conditions} of the unit cell.  
Impressive results are obtained for GaAs calculations using this approach on DFT+FEM that systematically 
converges, as the number of elements is increased, to the values obtained using the conventional plane wave calculations, 
employing Fourier transforms and Ewald summations.

The scope of the dimension attainable with FEM has also been demonstrated by Sterne {\it et al} \cite{Sterne}.
They have employed FEM to compute positron distributions and
lifetimes for various defect structures. Because the distribution is non-electron-like 
(concentrating away from the nuclei rather than around them), the generality of the FEM 
basis makes it particularly well suited for such problems. In order to study defects they have modeled a 5488 atom cell; 
as far as we know, this is one of the largest such calculations reported by any method to date. 

Another applications of FEM to large-scale \emph{ab initio} calculations is the work of Tsuchida {\it et al}
on molecular dynamics simulations of liquids \cite{TT}. They apply FEM study of 
liquid formamide (HCONH2) using a generalized gradient approximation to the exchange and correlation terms to improve the
modeling of hydrogen bonds. The use of adaptive coordinate transformations helped to double the resolution of the basis at the C, N and O centers. 
The liquid was modeled by 100 molecules in a cubic supercell of side 35.5 a.u., for a total of 600 atoms. 
The scope of this simulation is of the order of the largest one performed by standard PW methods 
to date and demonstrates the capacity of FEM for such large-scale computations already.

A new promising approach developed recently by Mazziotti \cite{Mazziotti1} is  the  large-scale 
semidefinite programming for many-electron quantum mechanics. In this approach, the energy of a many-electron quantum system can be approximated by a
 constrained optimization of the two-electron reduced density matrix (2-RDM) that is solvable in polynomial time by 
semidefinite programming (SDP).  The developed   SDP method for computing strongly correlated 2-RDMs  is 10-20 times faster 
than previous methods \cite{Mazziotti2}. This approach was illustrated for metal-to-insulator transition of H$_{50}$, with 
about $10^7$ variables.  

These examples along with our current work, suggest that we can combine the Finite-element method with finite size scaling 
to obtain critical parameters  in extended systems. One possibility is to calculate 
the phase transition at zero temperature in the 2-dimensional electron gas. 
The ground state energy of the system favors a crystallize state as the density of  
the electrons is lowered. Tanatar and Ceperey \cite{TC} performed a variational 
Monte Carlo simulations to the total energy using Hartree-Fock approach
\begin{equation}
  E_0(r_s) = E_{HF}(r_s) + E_c(r_s)
\end{equation}
They find the gas-to-crystal transition at $r_s = 37 \pm 5$. This system 
would be ideal to test FEM with finite size scaling for quantum criticality in extended systems.

\section{Conclusion}

The theory of finite size scaling (FSS) has been demonstrated to be of great utility in 
analyzing numerical data of finite systems \cite{fisher,widom,barber,privman,cardy,nightingale1,Peter1,Peter2}.
Moreover, it has been demonstrated that the theory also works for quantum systems, where the formalism applies to 
a complete basis expansion in the Hilbert space \cite{neirotti0,serra2,kais,snk1,snk2,nsk,qicun,kais1,review,adv}.
In this study we implemented the finite-element method which uses a local basis expansion 
consisting of arbitrary polynomials in real space and with variable resolution. 
It was demonstrated that this method can be used to compute matrix elements $\langle O \rangle^{(N)}$
(used in the finite size scaling formalism) accurately and with monotonic convergence with respect 
to the system size N. 
We have implemented the combined approach on different approximations 
to electronic calculations such as Hartree-Fock approximation (+ electron-electron correlation) 
and density-functional theory (DFT) under the local density approximation and found 
that even at the simplest level in DFT, the results we have obtained are in good agreement 
with other more accurate estimates such as large order perturbation theory \cite{Baker}. 

The finite-element method has already shown great promise for extended systems since it uses
local basis expansion to yield sparse matrices that lend the methodology easily to parallel architectures \cite{pask,AHM}. 
Furtermore, the Khon-Sham equations in DFT provides a general method for extended system. Therefore 
we expect this combined approach can be applied to electron structure in atoms, molecules, and extended systems
without loss of generality. Nevertheless, more studies need to be done to confirm the applicability of FSS 
with FEM for larger systems, where this combined approach can be used to 
shed some light into the nature of phase transitions.

\section*{Acknowledgments}
We would like to acknowledge the financial support of ARO and 
B. W.-K. thankfully acknowledges support form the NSF under the grant PHY-0969689.
We would also like to thank the referee for his helpful feedback and for pointing out
the form for the convergence error in FEM.

\end{document}